\documentclass[letterpaper,aps,prl,twocolumn,showcaps,superscriptaddress,showpacs,amsmath,amssymb]{revtex4-1}
\usepackage{color}
\usepackage{soul}
\usepackage{graphicx}
\usepackage{dcolumn}
\usepackage{bm}
\usepackage{units}
\usepackage{gensymb}

\renewcommand{\vec}[1]{\boldsymbol{#1}}

\graphicspath{{Figures/}}

\begin{document}

\preprint{APS/123-QED}
\title{Microscale Marangoni Surfers}
\author{Kilian Dietrich}
\thanks{These two authors contributed equally to this work. Author 
contributions are defined based on the CRediT (Contributor Roles Taxonomy) and 
listed alphabetically. Conceptualization: IB, KD, LI. Formal analysis: KD, LI, 
NJ, GV. Funding acquisition: IB, LI. Investigation: KD, NJ, GV. Methodology: 
IB, KD, LI, NJ, GV. Project administration: LI. Supervision: IB, LI. 
Validation: KD, NJ. Visualization: KD, LI, NJ. Writing - original draft: KD, 
LI, NJ. Writing - review and editing: IB, KD, LI, NJ, GV.}
\affiliation{Laboratory for Soft Materials and Interfaces, Department of Materials, ETH Z\"{u}rich, Z\"{u}rich, Switzerland.}
\author{Nick Jaensson}
\thanks{These two authors contributed equally to this work. Author 
contributions are defined based on the CRediT (Contributor Roles Taxonomy) and 
listed alphabetically. Conceptualization: IB, KD, LI. Formal analysis: KD, LI, 
NJ, GV. Funding acquisition: IB, LI. Investigation: KD, NJ, GV. Methodology: 
IB, KD, LI, NJ, GV. Project administration: LI. Supervision: IB, LI. 
Validation: KD, NJ. Visualization: KD, LI, NJ. Writing - original draft: KD, 
LI, NJ. Writing - review and editing: IB, KD, LI, NJ, GV.}
\affiliation{Laboratory for Soft Materials, Department of Materials, ETH Z\"{u}rich, Z\"{u}rich, Switzerland.}
\email{n.jaensson@gmail.com}
\author{Ivo Buttinoni}
\affiliation{Institut f\"{u}r Experimentelle Kolloidphysik, Heinrich-Heine Universit\"{a}t D\"{u}sseldorf, D-40225 D\"{u}sseldorf, Germany.}
\author{Giorgio Volpe}
\affiliation{Department of Chemistry, University College London, 20 Gordon Street, London WC1H 0AJ, United Kingdom.}
\author{Lucio Isa}
\email{lucio.isa@mat.ethz.ch}\affiliation{Laboratory for Soft Materials and Interfaces, Department of Materials, ETH Z\"{u}rich, Z\"{u}rich, Switzerland.}

\date{\today}

\begin{abstract}
We apply laser light to induce the asymmetric heating of Janus colloids adsorbed at water-oil interfaces and realize active micrometric "Marangoni surfers". The coupling of temperature and surfactant concentration gradients generates Marangoni stresses leading to self-propulsion. Particle velocities span four orders of magnitude, from microns/s to cm/s, depending on laser power and surfactant concentration. Experiments are rationalized by finite elements simulations, defining different propulsion regimes relative to the magnitude of the thermal and solutal Marangoni stress components. 
\end{abstract}

\maketitle

Microscale active materials constituted by ensembles of self-propelling colloidal particles offer tremendous opportunities for fundamental studies on systems far from equilibrium and for the development of disruptive technologies~\cite{bechinger2016active}. Central to their functions is the ability to convert uniformly distributed sources of energy, i.e.~under the form of chemical fuel or external driving fields, into net motion thanks to built-in asymmetry in their geometry or composition. Both from a modeling and a control perspective, minimalistic designs are particularly appealing. In such designs, the complexity of the particles is kept to a minimum, while still enabling functionality and the emergence of novel physical behaviors. The simplest case is the one of active Janus microspheres, i.e.~colloidal beads equipped with a surface patch of a different material, which exploit their broken symmetry to self-propel and yet reveal a broad range of complex phenomena, including dynamic clustering~\cite{Buttinoni_2013,Palacci_2013}, swarming~\cite{yan2016} and guided motion~\cite{Simmchen_2016}.

Self-motility in Janus particles can derive from various mechanisms, from catalytic reactions~\cite{sanchez2015chemically} to bubble propulsion~\cite{Wang_2017} and electrokinetic effects~\cite{Gangwal_2008}.  
Among the available propulsion schemes, self-phoretic mechanisms have emerged as a standard~\cite{moran2017phoretic}. In self-phoresis, a particle propels with a velocity that is proportional to a self-generated gradient via a phoretic mobility coefficient~\cite{anderson1989transport}. 
Self-thermophoresis, whereby motion is induced by the asymmetric heating of light-absorbing Janus particles~\cite{jiang2010active}, is particularly interesting due to the unique properties of light as a source of self-propulsion~\cite{Zemanek_19}. Here, the propulsion velocity is $\boldsymbol{V} = - D_T \mathbf{\nabla} T$, where both the thermophoretic mobility $D_T$~\cite{Braibanti_2008} and the self-generated thermal gradient are independent of particle size~\cite{Bregulla_2015}. The propulsion speed simply scales with incident illumination in a linear fashion, enabling robust possibilities for spatial and temporal motion control by light modulation~\cite{ilic2016guiding,Qian_2013,bregulla2014stochastic}. 

However, direct self-thermophoresis in bulk liquids is not an efficient propulsion mechanism. Its  molecular origin stems from particle-solvent interactions and the magnitude of $D_T$ is set by the temperature dependence of the particle(solid)-liquid interfacial energy \cite{parola2004particle}. Because the latter quantity depends weakly on temperature, thermophoretic mobilities are small. Typical values of $D_T$ are $ \mathcal{O}(\unit[ 10^{-12}]{m^2/s})$, leading to speeds of $\mathcal{O}(\unit[]{\mu m/s})$, i.e. just a few body-lengths per second for micrometric colloids, for an increase of the cap temperature ($\Delta T$) of $\unit[1]{K}$, as confirmed by experiments~\cite{jiang2010active, Bregulla_2015, bregulla2014stochastic} and theoretical predictions~\cite{Bickel2013}. A powerful way to improve the efficiency of thermal gradients for self-propulsion relies on coupling them to other gradients, such as asymmetric chemical gradients~\cite{buttinoni2012active, Gomez-Solano_2017}. 

In this Letter, by employing Janus particles at fluid-fluid interfaces, we couple thermal gradients to gradients of interfacial tension. Upon heating, controlled surface tension differences across the particle can lead to self-propulsion velocities up to staggering $10^4$ body-lengths per second, a vast increase over direct self-thermophoresis in bulk. This enhancement follows the fact that, in the presence of surface tension gradients, momentum conservation at the interface prescribes the existence of tangential stresses, called Marangoni stresses, defined as:
\begin{equation}\label{eq:Marangoni}
\nabla_\text{s} \sigma(\Gamma,T) = \frac{\partial \sigma(\Gamma, T)}{\partial T} \nabla_\text{s}T+\frac{\partial \sigma(\Gamma, T)}{\partial \Gamma} \nabla_\text{s}\Gamma,
\end{equation}
where  $\sigma(\Gamma,T)$ is the interfacial tension, which 
is a function of temperature $T$ and surface excess concentration of a surface-active species $\Gamma$. 
$\nabla_\text{s}=(\vec I - \vec n \vec n) \cdot \nabla$ is the surface gradient operator, with $\vec I$ the unit tensor and $\vec n$ the normal to the interface. Here, we identify two sources of stress: temperature  and surface excess concentration gradients, whose magnitude is set by $\partial \sigma(\Gamma, T)/\partial T$ and $\partial \sigma(\Gamma, T)/\partial \Gamma$, respectively \cite{RN990}. Imposing a force balance on the particle's surface $\partial P$ and contact line $L$ yields
\begin{equation}
\int_{\partial P} \boldsymbol{\sigma} \cdot \boldsymbol{n}_\text{p} \; dS= 
\int_{L} \sigma(\Gamma, T) \boldsymbol{t} \; dl, \label{eq:force_bal}
\end{equation}
where $\boldsymbol{\sigma}$ is the bulk stress tensor, $\boldsymbol{n}_\text{p}$ is the unit vector normal to the particle surface and $\vec t$ is the unit vector tangential to the interface and normal to the contact line.
Together with a no-slip boundary condition at the particle surface, Eq.~\eqref{eq:force_bal} allows solving for the particle velocity $\boldsymbol{V}$ imposed by the Marangoni stress~\cite{SupplementalMaterial}. For a characteristic interfacial tension difference $\Delta \sigma$,
simple dimensional arguments lead to predicting a propulsion speed $V \propto \Delta \sigma /\eta$, where $\eta$ is an effective viscosity experienced by the particle straddling the interface. Considering thermal Marangoni effects (aka thermocapillarity) alone, the predicted self-propulsion speed is given by $V \approx (\partial \sigma/\partial T) \Delta T /(10\eta) $~\cite{wurger2014thermally}. Typical values of $\partial \sigma/\partial T$ for oil-water interfaces are $ \mathcal{O}(\unit[ 10^{-4}]{N/(m\cdot K)})$, leading to speeds $ V= \mathcal{O}(\unit[]{cm/s})$, independent of particle size and indeed corresponding to $10^4$ body-lengths per second for microparticles, for $\Delta T= \unit[1]{K}$ and $ \eta = \mathcal{O}(\unit[10^{-3}]{Pa\cdot s})$~\cite{wurger2014thermally}.  The magnitude of $\partial \sigma/\partial T$ is then able to set macroscopic objects in motion, as for instance shown for the propulsion of centimeter-sized objects~\cite{Okawa2009} and the rotation of micro-gears suspended at a water-air interface~\cite{maggi2015thermocapillary}. Similar considerations can be made for solutal Marangoni propulsion~\cite{Lauga2012,Masoud2014}, as popularized by camphor or soap ``boats'' releasing surfactant at one end~\cite{strutt_1890,sur_2019}, or for the motion of active droplets~\cite{maass2016swimming}. 

\begin{figure}[t]
	\centering
	\includegraphics[width=0.44\textwidth,clip]{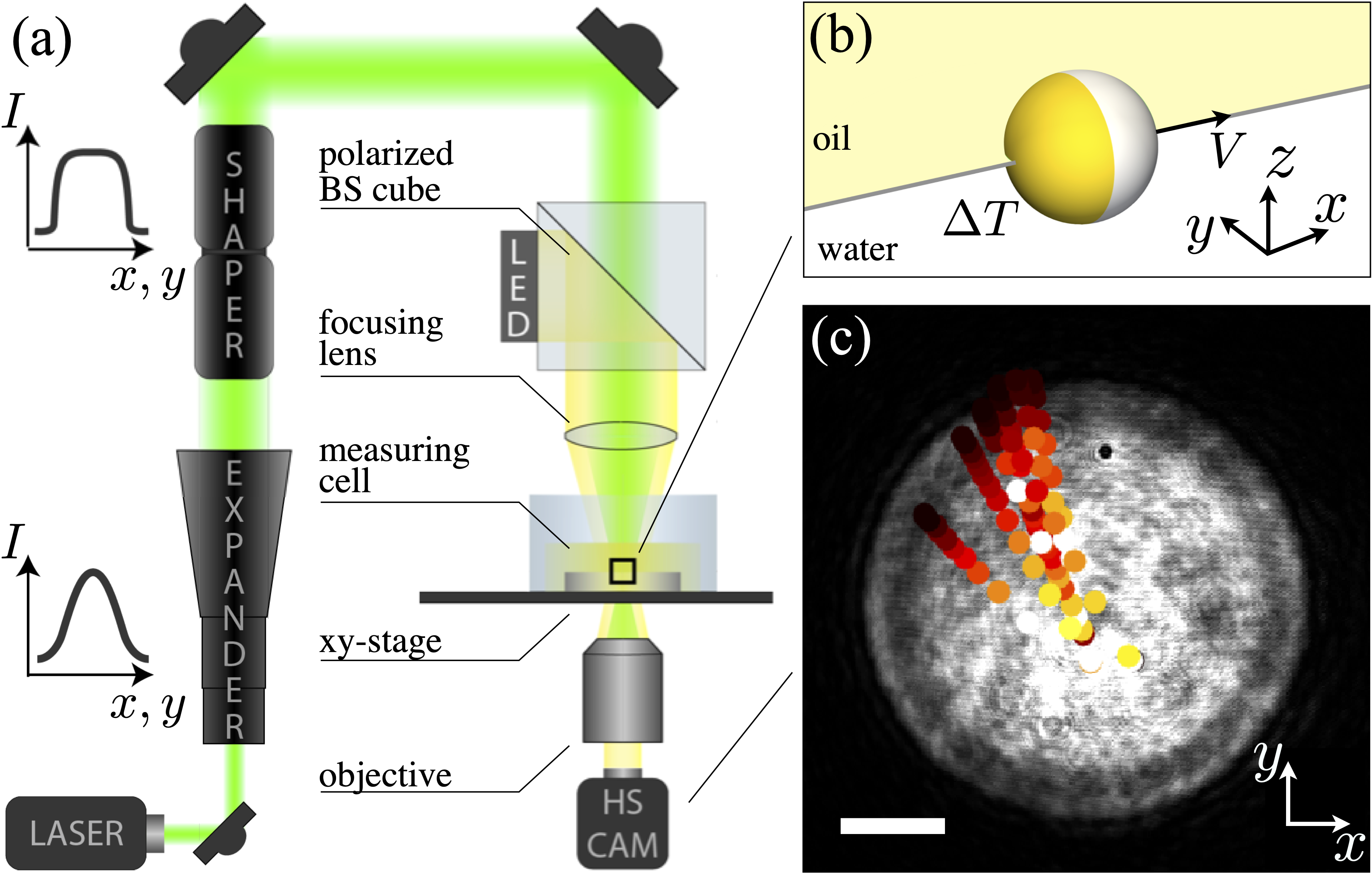}
	\caption{(a) Diagram of the experimental setup. 
	(b) Schematic of a gold-coated Janus particle at the water-oil interface. Laser illumination induces a temperature increase $\Delta T$ in the fluid in contact with the cap, leading to asymmetric Marangoni stresses and particle propulsion with velocity $V$. (c) Sample trajectories of Marangoni surfers propelling from the center of the illuminated region. The color code indicates normalized velocity ranging between zero (black) and a power-density-dependent $V_\text{max}$ (white). Scale bar = $\unit[30]{\mu m}$.}
	\label{fig:setup}
\end{figure}

Our interfacial microswimmers, or ``Marangoni surfers'', are Janus silica microparticles (radius $R=\unit[3.15]{\mu m}$, Microparticles GmbH) sputter-coated with a $\unit[100]{nm}$-thick hemisphere of gold (CCU-010, safematic) and confined at an interface between MilliQ water and dodecane (Arcos Organics, three-times purified through a basic alumina column). By pinning the lower aqueous phase to the sharp edges of a metal ring before adding a layer of dodecane, we achieve a macroscopically flat interface (area $\approx \unit[0.8]{cm^2}$) to which the particles are added via contacting the interface with a $\unit[0.5]{\mu l}$ droplet of a diluted aqueous suspension (0.01 \% w/v). The surface heterogeneity generated by the thick metallic caps effectively pins the Janus particles in random orientations with respect to the interface~\cite{Adams_2008,wang2016}, leading to caps typically crossing the interface (Figure S1 in~\cite{SupplementalMaterial}). Asymmetric heating of the particles is achieved by illuminating them with green laser light (2W-CW, Coherent Verdi, $\unit[532]{nm}$). In particular, we use beam-shaping optics to transform a Gaussian laser profile into a top-hat profile (Fig.~\ref{fig:setup}(a)) focused onto the interface plane to provide localized, spatially uniform illumination with a power density up to $\unit[8000]{W/cm^2}$~\cite{SupplementalMaterial}. Light absorption by the gold cap creates an asymmetric temperature profile around the particles~\cite{SupplementalMaterial}, thus generating Marangoni stresses that propel them with velocity $V$ and the Au cap oriented towards the back (Fig.~\ref{fig:setup}(b))~\cite{Bickel2013}. Trajectories are collected by positioning a particle in the center of the illuminated circular spot, turning the laser on at a given power density and recording images with a high-speed camera (AX 200 mini, Photron, up to 5000 fps) in a custom-built transmission microscope (Fig.~\ref{fig:setup}(c)). From the high-speed time lapses, we extract particle coordinates and velocities via Matlab particle-tracking algorithms. Particles self-propel from the center towards the periphery of the light spot in random directions, depending on the cap orientation (Movie S1). As soon as the particles leave the laser spot, propulsion stops (Movie S2). Particle speed, as a function of laser illumination, reaches values up to $\unit[1]{cm/s}$.



In order to rationalize the phenomenology seen in the experiments, we perform systematic numerical simulations of the fluid dynamics of the system coupled to heat and mass transport of surfactants at the interface using an in-house finite element code (full details in \cite{SupplementalMaterial}). 
In particular, as an ansatz to quantify the Marangoni stresses introduced in Eq.~\eqref{eq:Marangoni}, we assume that, in first approximation, the surface tension can be described by a linear function of  $\Gamma$ and $T$ \cite{Homsy1984}:
\begin{equation} 
\sigma(\Gamma,T) = \sigma_\text{0} - \delta \Gamma - \beta (T-T_0), 
\end{equation}
where $\sigma_0$ is the surface tension of the clean interface at the ambient temperature $T_0$.  $\delta$ and $\beta$ are the previously introduced parameters describing how the surface tension changes with $\Gamma$ and $T$, respectively identified with $-\partial \sigma / \partial \Gamma$ and $-\partial \sigma / \partial T$. Simulations were performed for Janus microparticles at a water-dodecane interface, where the model parameters were either known or taken from the literature \cite{SupplementalMaterial}. The results include the temperature, surface excess concentration and velocity fields as a function of $\Delta T$, which is the difference between the temperature of fluids in contact with the cap and $T_0$, and the equilibrium surfactant concentration $\Gamma_0$. 

\begin{figure}[hbt]
	\centering
	\includegraphics[width=0.44\textwidth,clip]{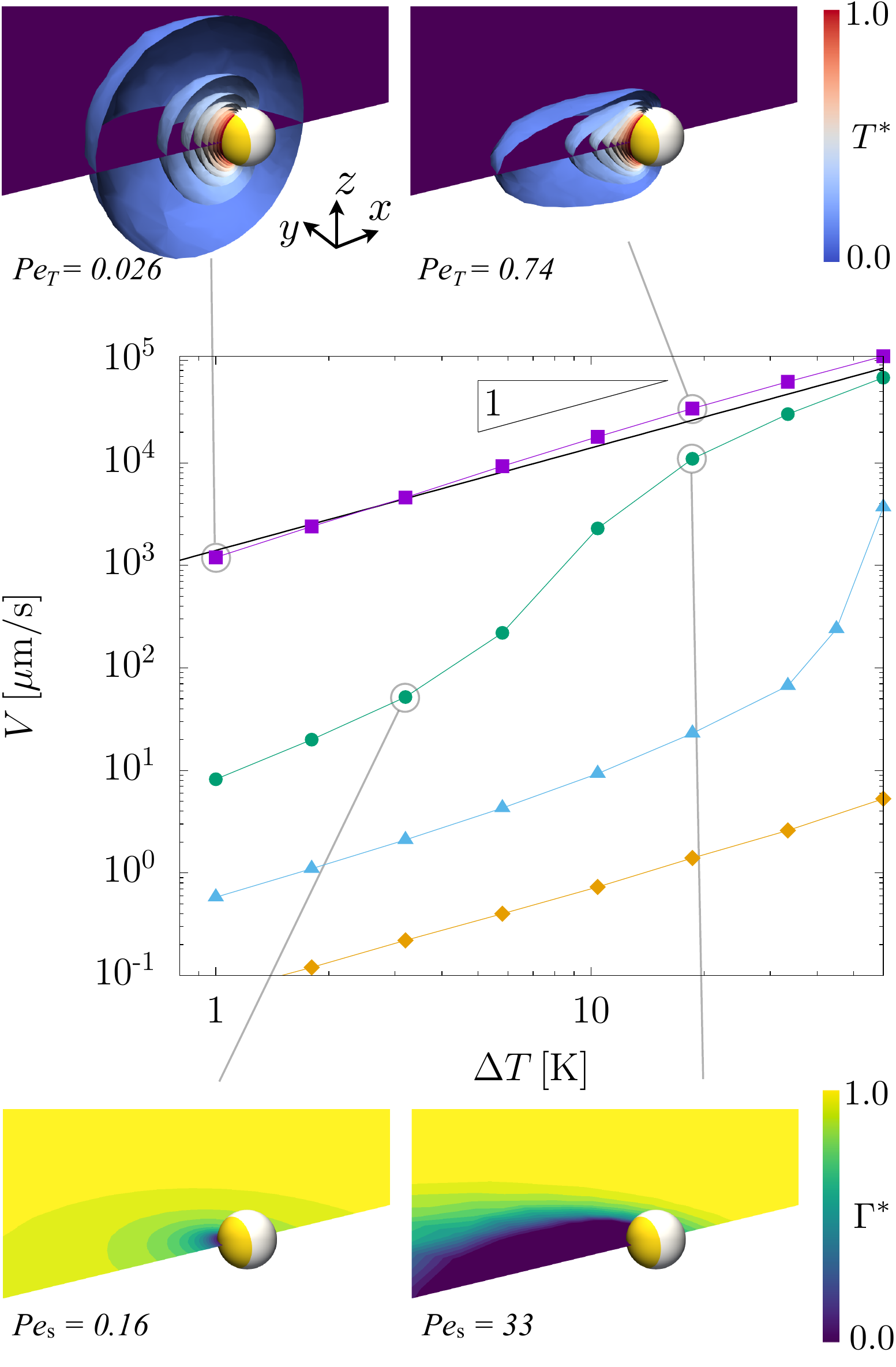}
	\caption{Simulated particle speed versus temperature increase $\Delta T$ in the fluid next to the cap of a Janus particle at a pristine water-dodecane interface (purple squares) and with excess concentrations  $\Gamma_0=10^{-7}$ (green circles), $10^{-6}$ (blue triangles) and $\unit[10^{-5}]{mole/m^2}$ (orange diamonds). The symbols are simulation results, and the connecting lines are to guide the eye. Solid black line: theoretical prediction by W\"{u}rger \cite{wurger2014thermally}. The insets show isosurfaces of the dimensionless temperature $T^* = (T-T_0)/\Delta T$ (top) and  excess concentration  $\Gamma^*=\Gamma/\Gamma_0$ (bottom) fields for selected simulations corresponding to different regimes of thermal and surface P\'{e}clet numbers  $Pe_\text{T}$ and $Pe_\text{s}$, respectively.}
	\label{fig:simulations}
\end{figure}

Starting from the case of a pristine interface ($\Gamma_0=0$), in Fig.~\ref{fig:simulations} we plot the simulated particle velocities $V$ as a function of $\Delta T$ (purple data). Here, we see that the speed increases roughly linearly with $\Delta T$. A dimensional analysis reveals that for fixed material parameters and in the absence of surfactants, the problem can be described by a single dimensionless group, for which we choose the thermal  P\'{e}clet number, defined by $Pe_T = 2VR/(\alpha_1+\alpha_2)$, where $\alpha_i$ are the thermal diffusivities of each liquid. Isosurfaces of the dimensionless temperature fields $T^* = (T-T_0)/\Delta T$ for the clean interface are shown in Fig.~\ref{fig:simulations} for $\Delta T = 1$ K (top, left) and 20 K (top, right), respectively corresponding to the case of $Pe_T <<1$ and $Pe_T \simeq 1$. 
For $Pe_T <<1$, the problem is governed by thermal diffusion and the temperature fields can be described as a combination of a monopole and a dipole solution. As the cap heating increases, the particle speed, and thus $Pe_T$, increase too, until convection starts affecting the temperature field, resulting in a stretched region of increased $T$ behind the particle. A direct comparison with existing theoretical analysis by W\"{u}rger
\cite{wurger2014thermally} shows very good agreement for the particle $V$ even in the high $Pe_T$ regime \cite{SupplementalMaterial}.

When we include the effect of surface-active species at the interface (green, blue and orange data), we obtain a markedly different behavior, which strongly depends on the value of $\Gamma_0$. We reveal the existence of a regime showing a linear velocity increase for low values of $\Delta T$, which expands over to broader ranges of $\Delta T$ for increasing surfactant surface excess, i.e. from green to orange data. However, the corresponding velocities are several orders of magnitude lower than the ones for the clean interface, even for a surface excess concentration as low as $\unit[10^{-7}]{mole/m^2}$. Interestingly, for this value of $\Gamma_0$ (green data), as $\Delta T$ grows, the particle speed goes through a transition regime and it converges towards the values for the clean interface. By increasing $\Gamma_0$, this transition takes place at correspondingly higher $\Delta T$. 

In order to understand this behavior, we visualize the surface excess concentration fields around the particles and revert once more to dimensional analysis. As compared to the clean interface, we need to introduce two additional dimensionless groups: a surface P\'{e}clet number, defined  $Pe_\text{s} = VR/D_\text{s}$, where $D_\text{s}$ is the surface diffusion coefficient, and the ratio between solutal and thermal Marangoni stresses, $\Pi=\beta \Delta T / (\delta \Gamma_0)$~\cite{Bickel2019}.
Since the surface diffusion coefficient can easily be two orders of magnitude lower than the thermal diffusion coefficient \cite{Wang_2013}, the transition $Pe_\text{s} \simeq 1$ happens at correspondingly lower $\Delta T$, for which the surface excess concentration field starts to deviate significantly from the diffusion-dominated regime.
The dimensionless surface excess concentration fields shown in Fig.~\ref{fig:simulations} (bottom) reveal that, as the particle moves, a wake depleted of surfactants is created behind it. Therefore, the corresponding surface concentration gradient generates solutal Marangoni stresses that ``work against'' the thermal Marangoni stresses. The resulting particle velocity is caused by the balance between the two stress components as in Eq.~\eqref{eq:Marangoni}, which explains the slowdown. 
The consequent transition to velocities purely dominated by thermal stresses happens because, at high $Pe_\text{s}$, the surface excess concentration inside the depleted region becomes approximately zero. At this point, the maximum attainable solutal Marangoni stress, fixed by $\delta \Gamma_0$, is reached and becomes independent of temperature, or particle speed, afterwards. Conversely, the maximum thermal Marangoni stress is set by $\beta \Delta T$, which keeps growing as the temperature increases. At a given point, thermal Marangoni stresess dominate, and the presence of the surfactant becomes insignificant, so that the data in Fig. 2 for a clean interface sets an upper limit for the particle velocity at a given $\Delta T$.

\begin{figure}[hbt]
	\centering
	\includegraphics[width=0.46\textwidth]{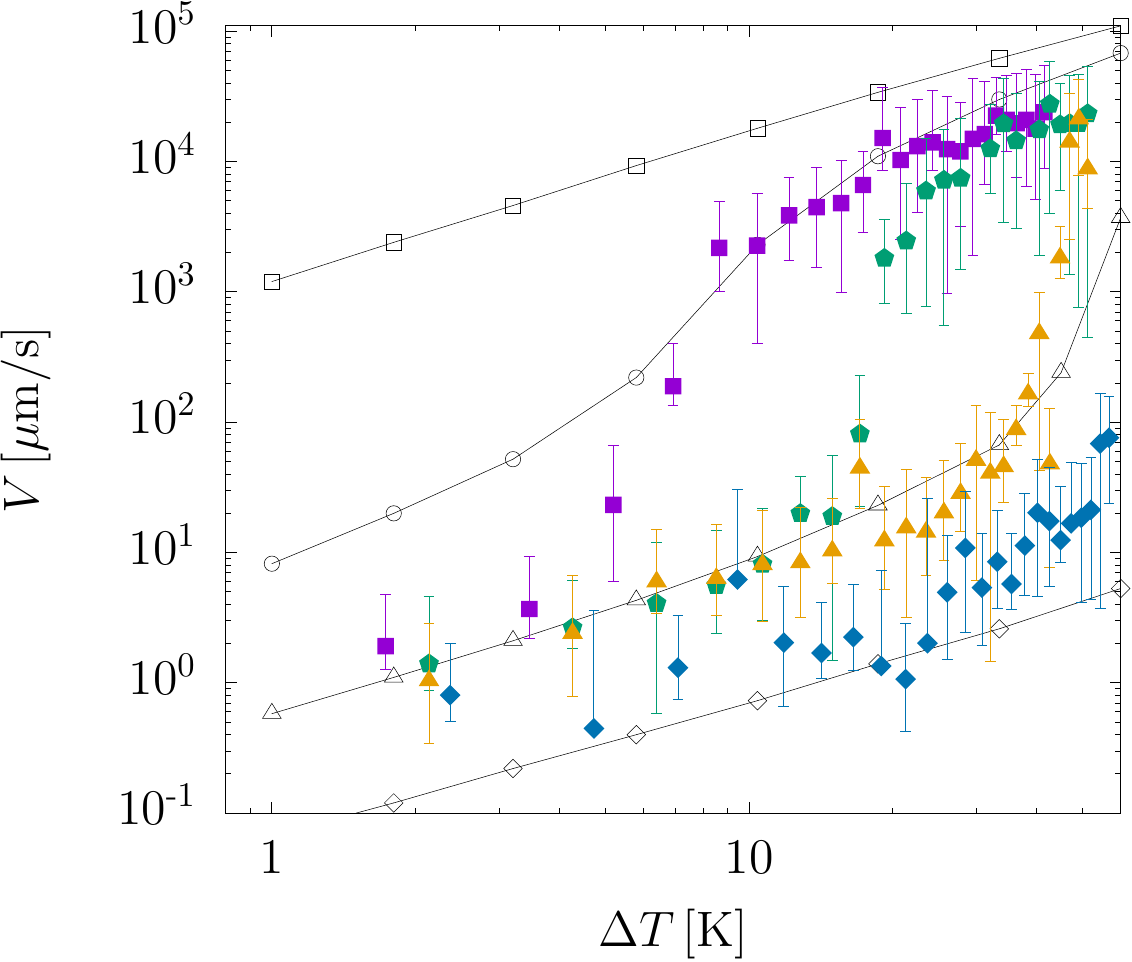}
	\caption{Experimental particle speed $V$ as a function of $\Delta T$. Allegedly pristine interface (purple squares) and increasing concentrations of SDS in the water phase: $C=10^{-7}$ (green pentagons), $10^{-5}$ (orange triangles) and $\unit[10^{-3}]{mole/L}$ (blue diamonds). Black, open symbols are simulated data with the symbol shape corresponding to Fig.~\ref{fig:simulations}.
	}
	\label{fig:experiments}
\end{figure}

Starting from these numerical predictions, we closely examine the measured experimental particle speeds as a function of incident laser power, which we convert into a $\Delta T$, leading to the data reported in Fig.~\ref{fig:experiments}. We perform the conversion by carefully calibrating the local heating of the fluids induced by the gold caps relative to the critical temperature of a water-lutidine mixture. We show a linear relation between the induced heating and incident laser power, as supported by theoretical estimates (see Supplemental Material \cite{SupplementalMaterial}). We first perform a series of measurements at an allegedly pristine water-dodecane interface. The purple data show a behavior consistent with the scenario reported by the simulations, where, in spite of all efforts for cleanliness, solutal effects are always present. For comparison, the simulation results are also plotted (empty symbols) and the experimental behavior can be reconciled by introducing a surface excess concentration of order $\Gamma_0=\unit[10^{-7}]{mole/m^2}$. These minute values of $\Gamma_0$ correspond to unavoidable environmental trace contaminations \cite{Liu2019,Maali2017,Manor2008}, which have a hardly measurable effect on the absolute level of the surface tension. However, as low as the absolute levels are, gradients of the surface excess concentration can still significantly alter the hydrodynamics in sensitive experiments, especially at small length scales, where Marangoni stresses become increasingly important \cite{Brennen2014}.

\begin{figure}[hbt]
	\centering
	\includegraphics[width=0.46\textwidth,clip]{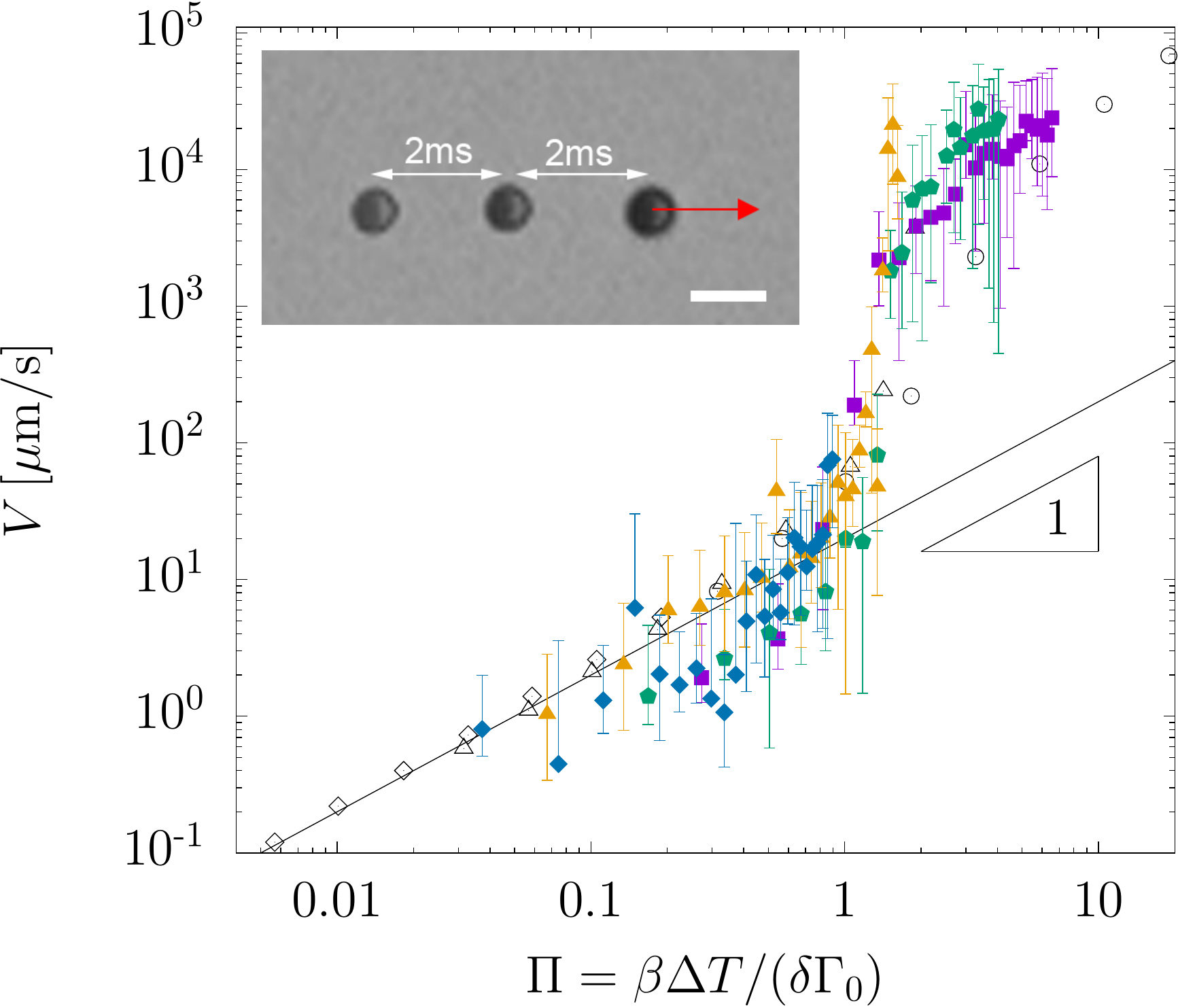}
	\caption{Rescaled experimental and simulation data. For the experiments the excess concentrations used are: $\Gamma_0=2 \times 10^{-7}$, $4 \times 10^{-7}$, $1 \times 10^{-6}$ and $\unit[2 \times 10^{-6}]{mole/m^2}$ for SDS concentrations of $C=0$, $ 10^{-7}$, $10^{-5}$ and $\unit[ 10^{-3}]{mole/L}$, respectively. The symbols correspond to the symbols in Fig.~\ref{fig:experiments}. The solid black line is a guide to the eye. (inset) Image overlay of a particle propelling with $V \simeq \unit[10]{mm/s}$ (scale bar = $\unit[10]{\mu 
	m}$).}
	\label{fig:rescaled}
\end{figure}

To confirm the role played by surface-active species, we purposely add controlled amounts of a water-soluble surfactant (sodium dodecyl sulfate, SDS, Sigma Aldritch, $\ge 98.5\%$). The choice of SDS is motivated by the fact that it has a negligible surface viscosity, and thus we do not expect surface rheology to affect particle motion \cite{Zell_2014}. Consistently with the balance between the different components of Marangoni stresses, we observe that an increased amount of SDS causes a shift of the transition towards higher values of $\Delta T$ and an overall reduced particle velocity over the same $\Delta T$ window. 
To unify the experimental and numerical data, and to unequivocally show that the ratio between the solutal and thermal Marangoni stresses controls the transition between the two regimes, we can rescale the experimental and simulation data on a master curve as a function of the dimensionless number $\Pi$ in Fig.~\ref{fig:rescaled}. Below $\Pi=1$, the data clearly collapses onto a single curve, with a transition that happens at $\Pi=1$ for all the data shown. We remark that the data only collapses for $\Pi<1$; for larger values of $\Pi$ the role of the solutal Marangoni stresses becomes insignificant and $V$ is rescaled by $\Delta T$ alone.

The understanding and rationalization of the experimental data opens up exciting opportunities for the exploitation of Marangoni stresses in self-propelled, active microscale systems. 
The green and purple data in Figs.~\ref{fig:experiments} and \ref{fig:rescaled} show that the propulsion velocity of a microscale Marangoni surfer can be tuned over four orders of magnitude in a single experiment via the controlled balance between thermal and solutal effects. In particular, the fact that this huge dynamic range can be regulated simply by light enables unprecedented opportunities for the spatial and temporal modulation of self-propulsion. The existence of two distinct linear regimes at low and high $\Pi$ allows fine-tuning of propulsion speed within two markedly different velocity ranges. At low $\Delta T$ velocities are on the order of $\unit[1-10]{\mu m/s}$, as typical for active Brownian particles. The simultaneous illumination of multiple particles with controlled light landscapes offers interesting options to modulate collective active motion. In the high $\Delta T$ regime, the particles move fully ballistically, with speeds reaching up to $\unit[20]{mm/s}$ (Inset of Figs.~\ref{fig:rescaled}), which had so far only been reported for bubble propulsion of microscale objects~\cite{sanchez2011jets}. Moreover, the narrow transition region together with the steep velocity variation can give rise to rich dynamical behavior crossing between regions of low and high $Pe \propto V/\sqrt{D_T D_R}$ for active motion, where $D_T$ and $D_R$ are the Brownian translational and rotational diffusivities. Finally, the strong dependence of propulsion speed on interface contamination may be used as a sensitive characterization tool for the presence of surface-active species undetectable by macroscopic tensiometry methods. 

In conclusion, from the demonstration of the first catalytically active particle onward~\cite{ismagilov2002autonomous}, fluid interfaces have been offering a broad range of promising opportunities to realize new active systems~\cite{Bishop2017}, exploiting the unique combination of strong vertical confinement~\cite{Dietrich_2017, Wang_2015}, specific interactions~\cite{Dietrich_2018} and highly efficient available propulsion sources. We expect that the near future will see further expansion, encompassing both fundamental studies and applications~\cite{Yao_2020}.\\

\acknowledgements{The authors thank M.A.~Hulsen at the Eindhoven University of Technology (TU/e) for access to the TFEM software libraries, and A. Studart, M. Fiebig and E. Dufresne at ETH Zurich and A. Fink at the Adolphe Merkle Institute for access to instrumentation. J. Vermant at ETH Zurich and R. Piazza at Politecnico di Milano are acknowledged for insightful comments. N.J. acknowledges TOTAL S.A.~for financial support. L.I and K.D acknowledge financial support from the ETH Research Grant ETH-16 15-1. L.I. and G.V. acknowledge financial support from the MCSA-ITN-ETN "ActiveMatter" 812780.} \\

\bibliography{refs}

\end{document}


\preprint{APS/123-QED}
\title{Microscale Marangoni Surfers: Supplemental Material}
\author{Kilian Dietrich}
\thanks{These two authors contributed equally to this work.}
\affiliation{Laboratory for Soft Materials and Interfaces, Department of Materials, ETH Z\"{u}rich, Z\"{u}rich, Switzerland.}
\author{Nick Jaensson}
\thanks{These two authors contributed equally to this work.}
\affiliation{Laboratory for Soft Materials, Department of Materials, ETH Z\"{u}rich, Z\"{u}rich, Switzerland.}
\email{n.jaensson@gmail.com}
\author{Ivo Buttinoni}
\affiliation{Institut f\"{u}r Experimentelle Kolloidphysik, Heinrich-Heine 
Universit\"{a}t D\"{u}sseldorf, D-40225 D\"{u}sseldorf, Germany.}
\author{Giorgio Volpe}
\affiliation{Department of Chemistry, University College London, 20 Gordon Street, London WC1H 0AJ, United Kingdom.}
\author{Lucio Isa}
\email{lucio.isa@mat.ethz.ch}\affiliation{Laboratory for Soft Materials and Interfaces, Department of Materials, ETH Z\"{u}rich, Z\"{u}rich, Switzerland.}

\maketitle
\beginsupplement
\section{Janus particle orientation at the interface}

\begin{figure}[ht]
	\begin{center}
		\includegraphics[width=0.7\textwidth]{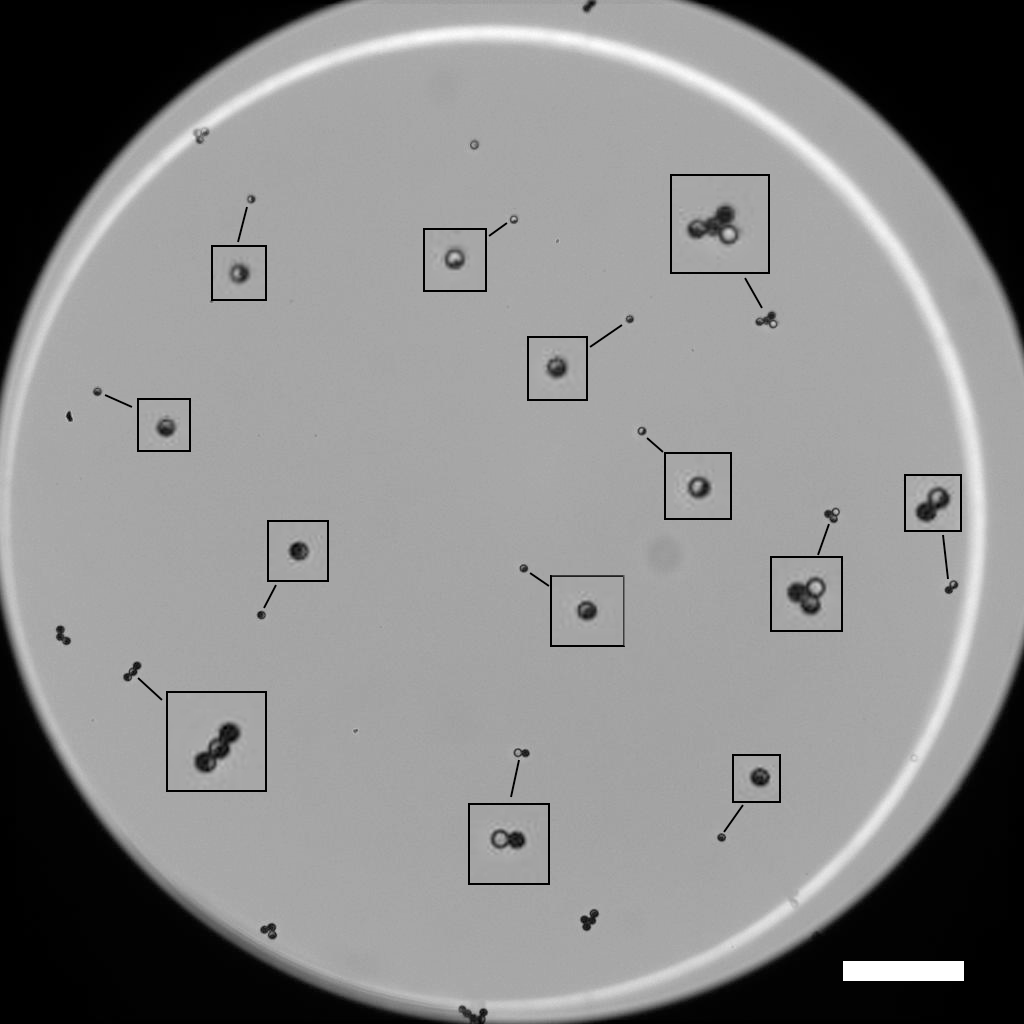}
	\end{center}
	\caption{Image of the silica Janus particles at the water-dodecane interface without laser illumination. Each box shows a zoomed-in image of the particles, demonstrating the presence of random orientations of the cap relative to the interface. The scale bar is $\unit[100]{\mu m}$.}
	\label{fig:orientation}
\end{figure}

\section{Measuring $\Delta T$}
\begin{figure}[hbt]
	\centering
	\includegraphics[width=0.95\textwidth]{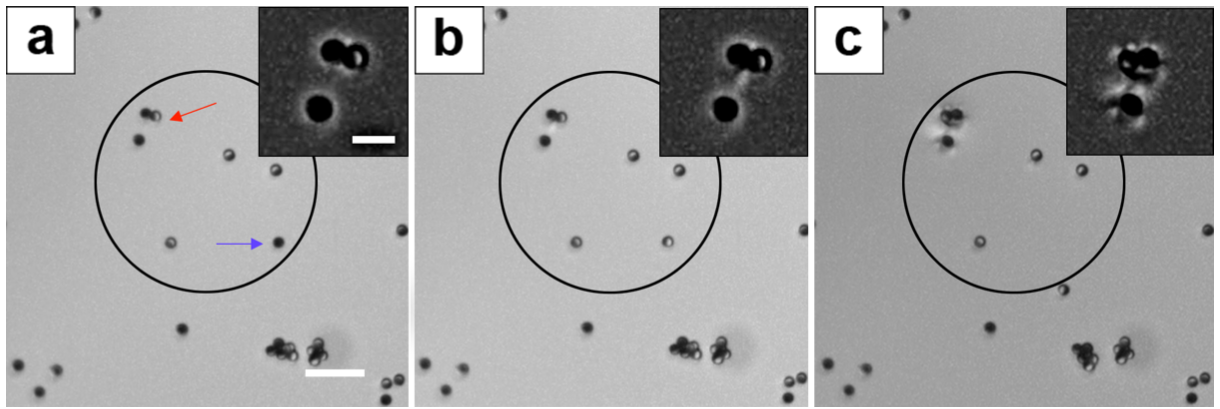}
	\caption{Demixing of a critical water-lutidine mixture in proximity of the caps of our Janus particles under laser illumination (spot corresponding to black circle). The bulk liquid temperature is $\unit[291]{K}$. (a) $I = \unit[32]{W/cm^2}$. (b) $I = \unit[2240]{W/cm^2}$. (c) $I = \unit[3200]{W/cm^2}$. The red arrow points to particles stuck to the substrate and the inset shows a zoomed-in image with enhanced contrast. The blue arrow points to a mobile particle. Demixing is observed in (b) corresponding to the caps reaching $T_\text{c}$ and causing mobile particles to start swimming within the perimeter of the laser spot. Upon exceeding $T_\text{c}$ (c) swimming and demixing are more pronounced.  The scale bars are $\unit[30]{\mu m}$ for (a,b,c) and $\unit[10]{\mu m}$ for all insets.}
	\label{fig:phase_sep}
\end{figure}

We calibrate the relationship between laser illumination power density $I$ and temperature increase $\Delta T$ for the fluid in the vicinity of the cap of our Janus particles by detecting the onset of the demixing of a critical mixture of DI water and lutidine (2,6-Dimethylpyridine) relative to a set temperature as a function of laser power. 
The critical mixture undergoes a phase transition when it exceeds its critical temperature $T_\text{c}$, corresponding to 307 K for a lutidine mass fraction of 0.286 in water, which we used.
The particles are dispersed in the mixture and injected into a quartz cell (Helma) sealed to avoid evaporation. The cell is placed on a temperature stage (Oko Labs), where temperature can be controlled from $\unit[273]{K}$ to $\unit[343]{K}$ with a precision of $\unit[0.1]{K}$. The temperature of the bulk liquid containing the particles is measured with a conductive wire inserted in the cell and the system is equilibrated for one hour at various temperatures between $\unit[281]{K}$ and $\unit[308]{K}$. These initial states correspond to well-defined $\Delta T$ relative to the demixing critical temperature. For every initial temperature, particles are illuminated under the same conditions as in the experiments and the laser power is increased in steps of 0.1 W until an effect can be clearly observed. Two phenomena can be detected, depending on whether the illuminated particles are stuck to the cell bottom or not. Mobile particles start swimming when demixing begins \cite{buttinoni2012active} and particles that are stuck on the glass substrate of the cell start demixing the surrounding liquid, as seen by a change in refractive index of the liquid surrounding the particle. The latter is perceived as a turbulent pattern, with a size of the order of the particle 
radius (shown in Fig.~\ref{fig:phase_sep}). Further increases in laser power lead to higher propulsion speed/stronger reactions around the particles that stop quickly after the laser is blocked or the power is reduced. 

\begin{figure}[hbt]
	\centering
	\includegraphics[width=0.5\textwidth]{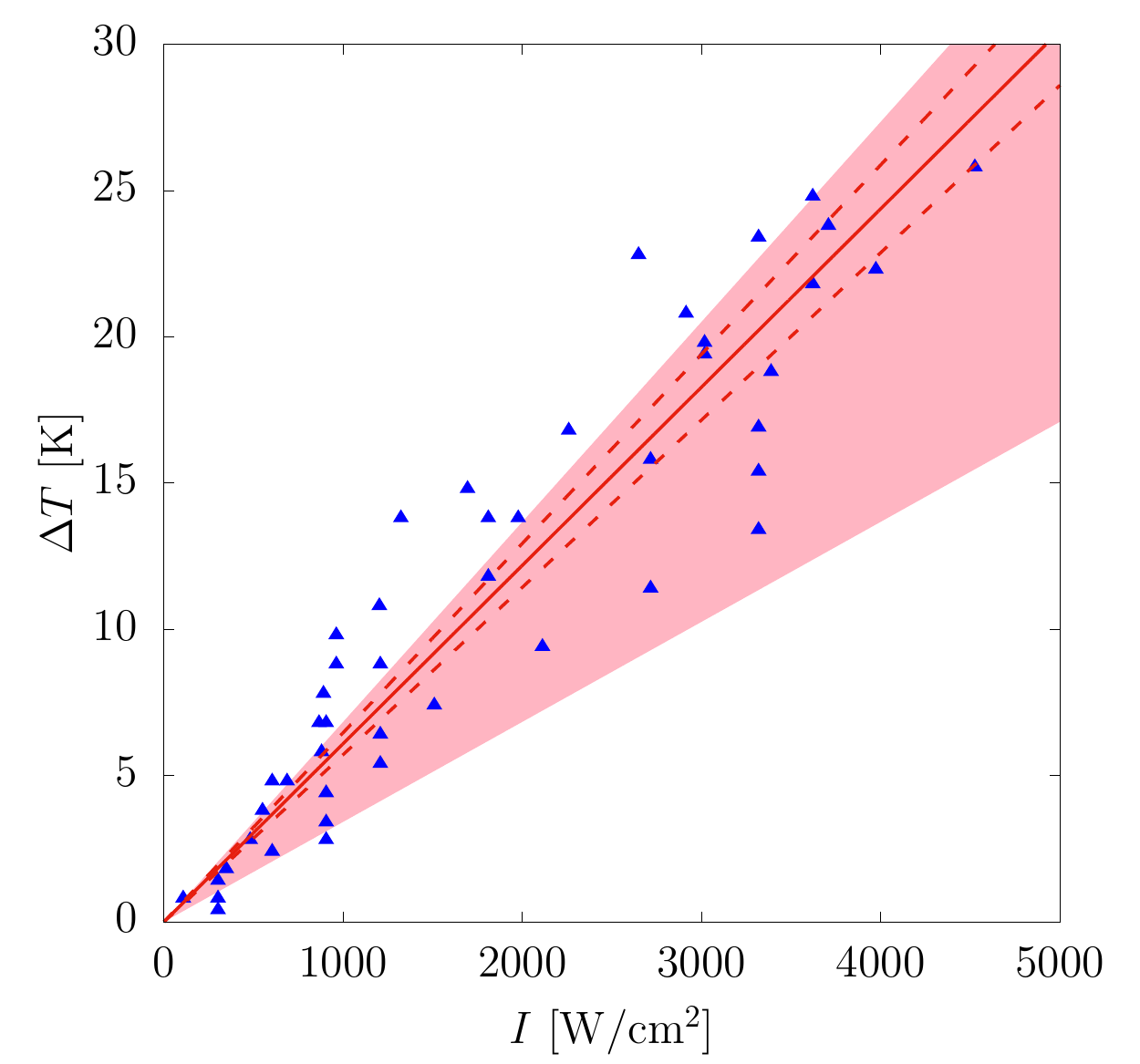}
	\caption{Calibration of fluid heating $\Delta T$ in the vicinity of the Au caps vs laser power density $I$. Blue triangles are data corresponding to different experiments. The solid red line is a linear fit to the data, with the dashed red lines representing 95\% confidence bands. The red-shaded area delimits the region between the upper and lower bounds of $\Delta T$ as estimated analytically (see text below).}
	\label{fig:temperature}
\end{figure}

A calibration of temperature increase vs laser power density relative to the background temperature allows us to plot particle velocities vs $\Delta T$, as shown in the main text. The calibration results are shown in 
Fig.~\ref{fig:temperature} and demonstrate a linear scaling between $\Delta T$ and laser power density.

\section{Analytical estimation of cap heating}\label{sec:analytical_temperature}

We support the calibration shown in Fig.~\ref{fig:temperature} by simple calculations. Here, as in the experiments, we consider a Janus sphere ($R = 3.15 \, {\rm \mu m}$) coated with gold on 25\% or 50\% of its surface. The thickness of the gold cap is $h = 100 \, {\rm nm}$. The surface is placed at the interface between water and dodecane ($\kappa_{\rm w} = 0.6 \, {\rm W K^{-1} m^{-1}}$ and $\kappa_{\rm d} = 0.14 \, {\rm W K^{-1} m^{-1}}$). The particle is illuminated by a laser beam at $532 \, {\rm nm}$ with an intensity of $I = 1 \, {\rm kW cm^{-2}}$. The temperature profile around a Janus particle can be derived from Fourier's law \cite{Bickel2013}
\begin{equation}
	\kappa\nabla^2T=q({\bf r})
\end{equation}
where $q$ is the power absorbed by the metal cap and $\kappa$ is the thermal conductivity of both the particle and the surrounding fluid. The estimations below neglect finite temperature jumps (Kapitza resistance) at the metal-fluid and metal-silica interfaces. Therefore, the temperature of the metal cap coincides with the one of the fluid at contact.

The cap conductivity is $\kappa_{\rm c} = 318  {\rm W K^{-1} m^{-1}}$. As in our case $\frac{\kappa_{\rm c}}{\kappa_{\rm d}} > \frac{\kappa_{\rm c}}{\kappa_{\rm w}} > \frac{R}{h}$, the cap forms an isotherm and we can assume a constant cap temperature $T_\infty + \Delta T$, where $T_\infty$ is the background temperature at infinite distance from the particle. If we consider the total heat flow from the particle as the power $P$ absorbed by the metal cap, we can obtain an expression for the excess temperature $\Delta T$ \cite{Bickel2013}
\begin{equation}
	\Delta T = \frac{P}{2 (\pi + 2)\kappa R}
\end{equation}
We can assume that $P = \epsilon I S$, where $\epsilon$ is the absorption efficiency of the metal cap and $S = 4 \pi \phi R^2$ is the surface of the metal cap with $\phi$ being the coverage factor over the sphere total surface. To take into account the liquid interface (dividing the cap in two exact halves), we can, in first approximation, estimate $\Delta T$ as the sum of the increase in temperature due to each of the two half-caps immersed in water and dodecane, respectively:
\begin{equation}
	\Delta T = \frac{\epsilon I S}{4 (\pi + 2)R} \frac{k_{\rm w}+k_{\rm d}}{k_{\rm w}k_{\rm d}} = \epsilon I \frac{ \pi \phi R}{(\pi + 2)} \frac{k_{\rm w}+k_{\rm d}}{k_{\rm w}k_{\rm d}}
\end{equation}

To get a range of possible values for $\Delta T$, we need to estimate $\epsilon$. 


An estimate for $\epsilon$ (and hence $\Delta T$) of our particles can be obtained based on previous experimental values of the temperature at the cap for slightly smaller Janus particles in water with thinner metal caps ($\epsilon = 0.0224$ for $R = 1.5 \, {\rm \mu m}$ and $h = 25 \, {\rm nm}$) \cite{Sano2010}. If we account for the fact that the skin depth of gold is $45 \, {\rm nm}$ \cite{Olmon2012}, our particles' absorption efficiency should be higher by a factor $\ell =$ 1.8 or 3.6 than those in \cite{Sano2010}, if we consider that radiation is absorbed by the first 45 or 90 nm of gold, respectively. In this more realistic estimation, $\Delta T = 1.71-3.42 \, {\rm K}$ for $\phi = 0.25$ and $\Delta T = 3.42-6.84 \, {\rm K}$ for $\phi = 0.5$. The upper and lower bounds for the case of $\phi = 0.5$ are reported in Fig.~\ref{fig:temperature}.

\section{Simulations}

We provide here the full details of the finite-elements numerical simulations of the fluid dynamics of the system coupled to heat and mass transport of surfactants at the interface of which we reported the salient results in the main manuscript.

\begin{figure}[ht]
	\begin{center}
		\includegraphics[width=0.6\textwidth]{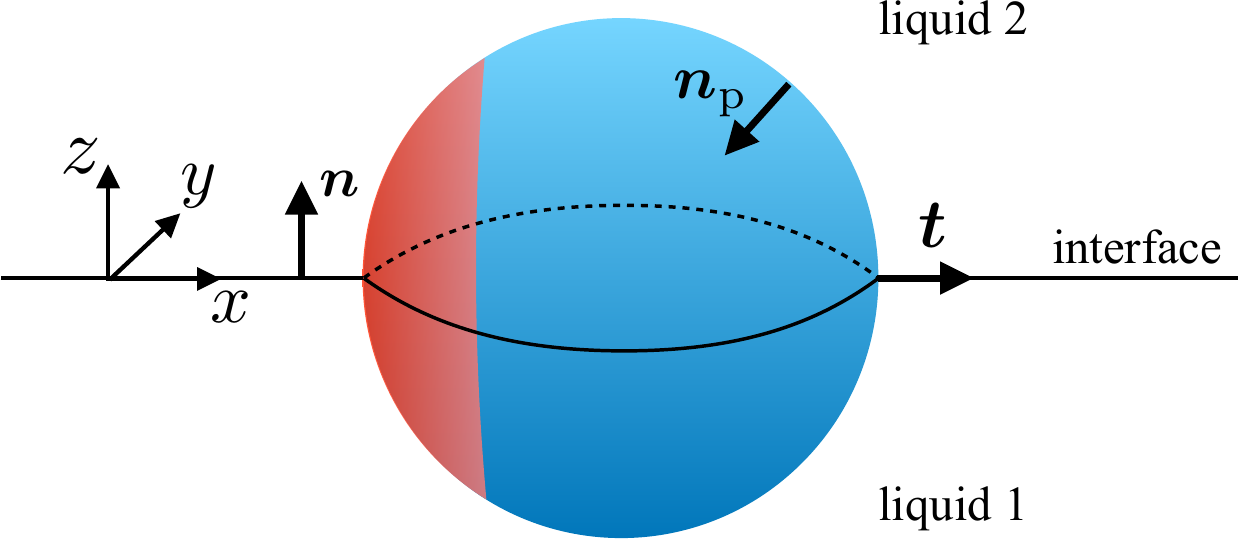}
	\end{center}
	\caption{Schematic of a Marangoni surfer at a liquid-liquid interface.}
	\label{fig:marangoni_surfer}
\end{figure}

We consider a spherical particle embedded in a liquid-liquid interface, which is endowed with a surface tension $\sigma(\Gamma,T)$, where $T$ is the temperature and $\Gamma$ is the surface excess concentration of a surfactant. A spherical cap located on the particle is heated to a temperature $T = T_0 + \Delta T$, where $T_0$ is the ambient temperature. This local increase in  temperature results in Marangoni flows and movement of the particle. The interface is located in the $xy$-plane, and is assumed to remain straight, i.e.~the capillary number $Ca = \eta_i v/\sigma << 1$, where $\eta_i$ and $v$ are the viscosities of each fluid and a characteristic fluid velocity, respectively. Moreover, for simplicity, the contact angle of the liquid-liquid interface with the particle boundary is assumed to be 90$^{\circ}$, i.e.~the plane in which the interface is located intersects the center point of the particle. Finally, we assume that the contact line is pinned, which implies that the particle does not rotate. A schematic of the problem is shown in Fig.~\ref{fig:marangoni_surfer}.

\subsection{Governing equations}
To describe the problem, we denote the domains occupied by the lower and upper liquids by $\Omega_1$ and $\Omega_2$, respectively. The boundary shared between $\Omega_1$ and $\Omega_2$ is denoted by $\partial \Omega$ and is identified with the liquid-liquid interface. The boundary of the particle is denoted by $\partial P$ and the line shared by $\partial \Omega$ and $\partial P$ is denoted by $L$, which is identified with the contact line of the liquid-liquid interface with the particle boundary. Moreover, the particle boundary is split into the heated cap $\partial P_\text{h}$ and the rest of the particle $\partial P_\text{c}$. The unit vectors $\vec n$, $\vec n_\text{p}$ and $\vec t$, are the normal to the interface, the normal to the particle boundary and the vector that is both normal to the contact line and tangential to the interface, respectively (see Fig.~\ref{fig:marangoni_surfer}). To write the components of these vectors, we use the Cartesian coordinate system as shown in Fig.~\ref{fig:marangoni_surfer}, e.g.~the normal vector is written as $\vec n = n_x \vec e_x + n_y \vec e_y+n_z \vec e_z$, where $\vec e_x$, $\vec e_y$ and $\vec e_z$ are the Cartesian basis vectors.

It is assumed that inertia plays no role and the liquids are incompressible, thus the flow in each domain is described by the Stokes equations, given by
\begin{align}
  -\nabla \cdot (2\eta_i \boldsymbol{D}) + \nabla p &= \boldsymbol{0} \\
    \nabla \cdot \vec v &= 0,
\end{align}
where $\vec v$ is the fluid velocity, $\eta_i$ is the viscosity of the $i$-th liquid, $\boldsymbol{D} = \left(\nabla\vec v + (\nabla \vec v)^T\right)/2$ is the rate-of-strain tensor and $p$ is the pressure. \\

The temperature in each domain is described by a convection-diffusion equation, which reads
\begin{equation}
\frac{\partial T}{\partial t}+\vec v \cdot\nabla T =\alpha_i \nabla^2T,
\label{eq:temp_cons}
\end{equation}
where $\alpha_i$ are the thermal diffusivities of each liquid, given by $\alpha = \kappa / (\rho c_p)$, where $\kappa$ is the thermal conductivity, $\rho$ is the density and $c_p$ is the specific heat at constant pressure.  \\

Concerning the transport of surfactants at the interface, we assume that the surfactant is insoluble and that there are no chemical reactions occurring, yielding the following balance equation for $\Gamma$ on $\partial \Omega$
\cite{Stone1990}:
\begin{equation}
\frac{\partial \Gamma}{\partial t} + \vec v \cdot \nabla_\text{s} 
\Gamma + (\nabla_\text{s} \cdot 
\vec v)\Gamma=D_\text{s}\nabla^2_\text{s} \Gamma,
\label{eq:surf_cons}
\end{equation}
where $\nabla_\text{s}=(\vec I - \vec n \vec n) \cdot \nabla$ is the surface gradient operator, with $\vec I$ the unit tensor, and $D_\text{s}$ is the surface diffusion coefficient. Note that we have used the assumption of the interface remaining straight in Eq.~\eqref{eq:surf_cons}.

Finally, we introduce a linear equation of state to relate the surface tension to the surfactant concentration and temperature 
\cite{Homsy1984}:
\begin{equation}
\sigma(\Gamma,T) = \sigma_0 - \delta \Gamma - \beta (T-T_0), \label{eq:surf_ten}
\end{equation}
where $\sigma_0$ is the surface tension of the clean interface at the ambient temperature $T_0$, and $\delta$ and $\beta$ are parameters that describe how the surface tension changes with $\Gamma$ and $T$, respectively. Note that the absolute value of the surface tension does not play a role due to the assumption of a straight interface, but is included here for completeness.

\subsection{Boundary and initial conditions}
The boundary conditions for the velocity are given by no-slip on the particle boundary, and no-fluid-flow far away from the particle:
\begin{alignat}{3}
  \vec v &= \vec V &\quad &\text{on } \partial P \label{eq:no_slip_part} \\
  \vec v &= \vec 0 &\quad &\text{for } |\vec x - \vec X| \rightarrow 
  \infty,  
\end{alignat}
where $\vec V$ is the (unknown) particle velocity, $\vec x$ is the position vector and $\vec X$ is the position vector of the center point of the particle. In order to solve for $\vec V$, a force balance on the particle must be satisfied, which is given by
\begin{equation}
\int_{\partial P} \boldsymbol{\sigma} \cdot \boldsymbol{n}_\text{p} \; dS= 
\int_{L} \sigma(T,\Gamma) \boldsymbol{t} \; dl. \label{eq:force_bal}
\end{equation}
where $\vec \sigma = - p \vec I + 2\eta_i \boldsymbol{D}$ is the Cauchy stress tensor.
At the interface between the two liquids, the velocity in the normal direction is zero, whereas no-slip is assumed for the velocities in the tangential direction:
\begin{alignat}{3}
  v_x|_1 &= v_x|_2 &\;\;\;\;\;\text{on } &\partial \Omega 
  \label{eq:no_slip_int1} \\
  v_y|_1 &= v_y|_2 &\;\;\;\;\;\text{on } &\partial \Omega 
  \label{eq:no_slip_int2} \\
  v_z|_1 &= v_z|_2 = 0 &\;\;\;\;\;\text{on } &\partial \Omega, 
\end{alignat}
where the notation $|_i$ implies that the variables is evaluated on the $i$-th side of the interface. Moreover, conservation of momentum leads to
\begin{equation}
(\vec \sigma|_1-\vec \sigma|_2) \cdot \vec n = \nabla_\text{s} \sigma \;\;\;\; 
\;\text{on } \partial \Omega. \label{eq:stress_bc}
\end{equation}
Note that we have used the assumption of the interface remaining straight in Eq.~\eqref{eq:stress_bc}. The right hand side of Eq.~\eqref{eq:stress_bc} can be expanded as
\begin{align}
\nabla_\text{s} \sigma = -\beta \nabla_\text{s}T-\delta \nabla_\text{s}\Gamma,
\label{eq:expansion}
\end{align}
where the first term on the right hand side is the \emph{thermal Marangoni stress} and the second term on the right hand side is the \emph{solutal Marangoni stress}.\\

For the temperature, the boundary conditions are given by prescribed temperatures for the fluid in contact with the heated cap and far away from the particle, and an insulating condition is used on the particle boundary that does not belong to the cap:
\begin{alignat}{3}
T &= T_0+ \Delta T &\quad &\text{on } \partial P_\text{h} \\
\vec n_\text{p} \cdot \nabla T  &= 0 &\quad &\text{on } \partial P_\text{c} 
\\
T &= T_0  &\quad &\text{for } |\vec x - \vec X| \rightarrow 
\infty.  
\end{alignat}

At the liquid-liquid interface, it is assumed that there is no thermal (Kapitza) resistance to heat flow, which yields continuity of the temperature field and heat flux:
\begin{alignat}{3}
T|_1 &= T|_2 &\quad &\text{on } \partial \Omega \label{eq:temp_bc1} \\
\vec n  \cdot \kappa_1 \nabla T|_1 &= \vec n  \cdot \kappa_2 \nabla  T|_2 
&\quad &\text{on } \partial \Omega. \label{eq:temp_bc2}
\end{alignat}
The temperature of the interface, which is needed to evaluate the surface tension in Eq.~\eqref{eq:surf_ten}, is thus equal to the temperature of the bulk, evaluated at the interface location.

For the surfactant, we assume a uniform surfactant concentration far away from the particle, and a no-flux condition at the contact line:
\begin{alignat}{3}
\vec t \cdot \nabla_\text{s} \Gamma  &= 0 &\quad &\text{on } L
\\
\Gamma &= \Gamma_0  &\quad &\text{for } |\vec x - \vec X| \rightarrow 
\infty.  
\end{alignat}
Initial conditions are needed for $T$ and $\Gamma$. It is assumed that the temperature is initially $T_0$ in the entire domain, except for the heated cap on the particle, which is set to $T_0+\Delta T$. For the surfactant, a uniform initial surfactant concentration, given by $\Gamma_0$, is used.

Finally, the particle velocity and position are related through the kinematic equation:
\begin{equation}
\vec V = \frac{d \vec X}{d t}.
\end{equation}

\subsection{Numerical method}
\begin{figure}[ht]
	\begin{center}
		\includegraphics[width=0.5\textwidth]{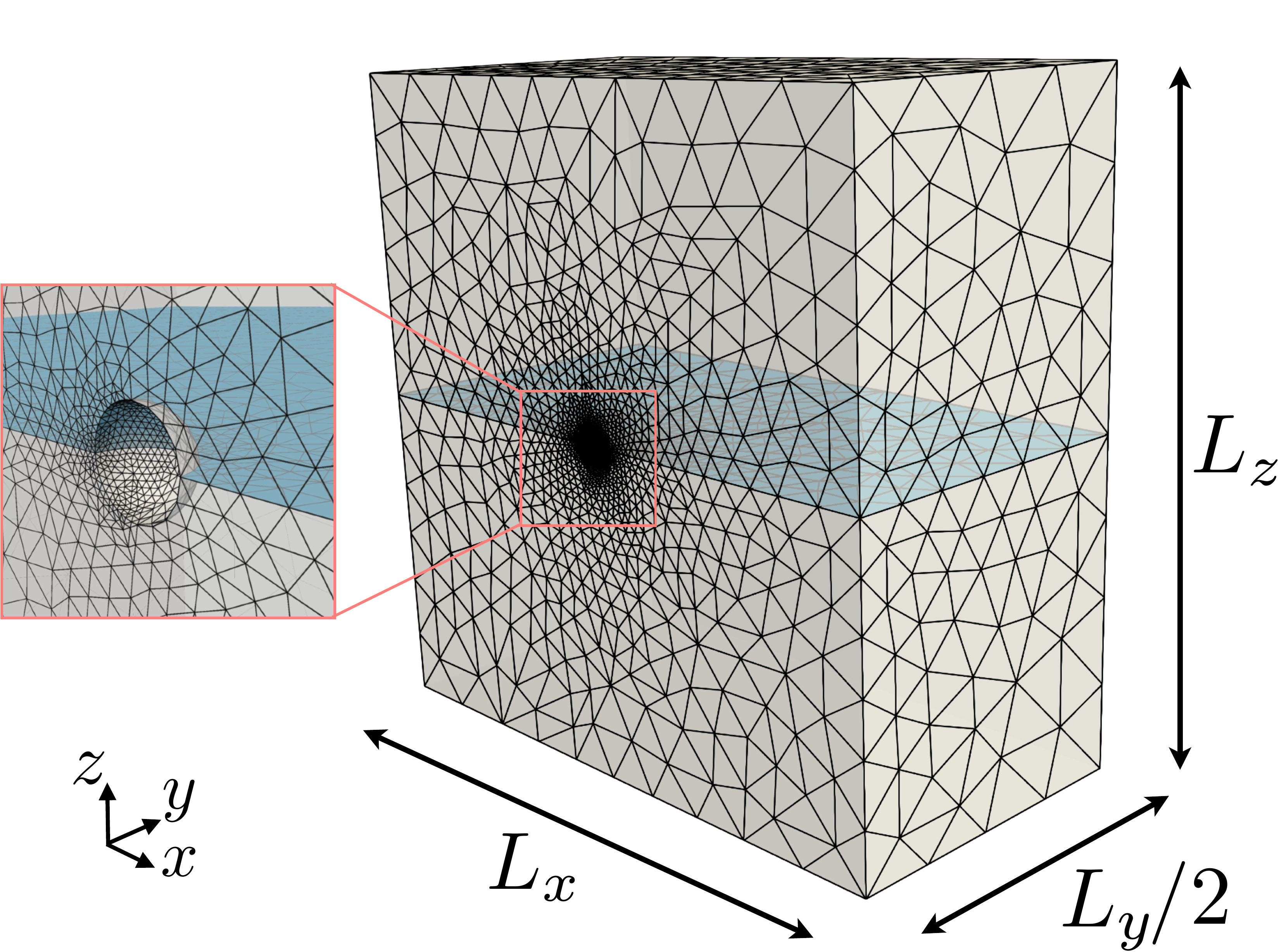}
	\end{center}
	\caption{Typical finite element mesh used in the simulations. The blue surface depicts the liquid-liquid interface.}
	\label{fig:mesh}
\end{figure}

The full system of coupled equations is solved using an in-house finite element code. In order to make the simulations more efficient, we exploit the symmetry of the problem in $xz$-plane, i.e.~only half of the domain is simulated and the particle can only move in $x$ direction ($\vec V = V \vec e_x$). The open-source mesh-generator Gmsh \cite{geuzaine2009} is used to generate tetrahedral meshes which are split into two parts, representing the two liquids (see Fig.~\ref{fig:mesh}). The element boundaries are aligned with the liquid-liquid interface and with the boundary of a particle embedded in the interface, which allows for a straightforward implementation of the boundary conditions. We use a large domain size so that boundary effects can be neglected: $L_x=L_y=L_z=100a$.

We use iso-parametric, tetrahedral P2/P1 (Taylor-Hood) elements for the velocity/pressure, and isoparametric, tetrahedral P2 elements for the temperature. The surface excess concentration is discretized on a surface mesh as shown in Fig.~\ref{fig:mesh}. For simulations at low surface P\'{e}clet number, $\Gamma$ is discretized using triangular P2 elements. For simulations at high surface P\'{e}clet number, triangular P1 elements are used, employing SUPG for stabilization \cite{Brooks1982}.

The boundary conditions in Eqs.~\eqref{eq:no_slip_part}, \eqref{eq:force_bal}, \eqref{eq:no_slip_int1}, \eqref{eq:no_slip_int2}, \eqref{eq:stress_bc}, \eqref{eq:temp_bc1} and \eqref{eq:temp_bc2} are implemented using constraints, yielding additional Lagrange multipliers in the system of unknowns. Moreover, the particle velocity $V$ is included in the system of unknowns, and its solution will be such that the force boundary condition Eq.~\eqref{eq:force_bal} is satisfied \cite{Glowinski1999}. The remaining boundary conditions prescribe known values, which can be filled in directly. More information on the numerical implementation for similar problems can be found in previously published work \cite{Jaensson2016,Jaensson2017}. 

For the transient results, second-order finite-differencing schemes are used to approximate the time derivatives. Within a time-step, the temperature equation is solved first using an estimate of the velocity from the previous time steps. Then, depending on the regime of the simulations, we employ one of two approaches: 1) the velocity/pressure and surfactant are solved in one system, using a Picard iteration for the non-linear terms, or 2) we first solve the surfactant equation and then the velocity/pressure system using the updated surfactant results. It was found that the former approach is crucial for stability in simulations that are dominated by surface elasticity (i.e.~high values of $\Gamma_0$), whereas the latter approach is crucial when the surface P\'{e}clet number is large. Finally, we note that the walls do not play a role in this problem, and we can therefore follow the particle motion by moving the whole mesh in $x$ direction. Using a Lagrange-Euler formulation, this implies that the particle velocity is subtracted from the convective terms in Eq.~\eqref{eq:temp_cons} and Eq.~\eqref{eq:surf_cons}, similar to  \cite{Villone2011}.

As we will show in the next section, the transients in the problem are fast, and the steady-state solutions suffice for a comparison to the experiments. Therefore, we also implemented the solution for the steady state directly by neglecting the $\partial / \partial t$ terms, and using a Picard iteration. This was found to speed up the calculations considerably, but does not always lead to converging solutions. For the solutions that do not converge, we use the transient simulations, and time-integration is performed until 
the solution has reached steady-state. All simulations were performed on the Euler cluster at ETH Z\"{u}rich.

\subsection{Results}
Simulations are performed for a particle of radius $a = 3.15 \times 10^{-6} \, \text{m}$ at the interface between water and dodecane. The bulk parameters used are the typical values for a water-dodecane system:
$\rho_1 = 997 \, \text{kg}/\text{m}^3$, 
$\rho_2 = 750 \, \text{kg}/\text{m}^3$, 
$\kappa_1 = 0.601 \, \text{W}/(\text{m K})$, 
$\kappa_2 = 0.14 \, \text{W}/(\text{m K})$, 
$(c_p)_1 = 4.18 \times 10^3 \, \text{J}/(\text{kg K})$, 
$(c_p)_2 = 2.21 \times 10^3 \, \text{J}/(\text{kg K})$, 
$\eta_1 = 0.89 \times 10^{-3} \, \text{Pa s}$,
$\eta_2 = 1.36 \times 10^{-3} \, \text{Pa s}$ \cite{CRC2019}.
All experiments are performed at room temperature, thus the ambient temperature $T_0 = 295 \, \text{K}$ and
the parameters for the interfacial tension are: $\sigma_0 = 52.6 \times 10^{-3} \, \text{N/m}$ and $\beta = 0.0757 \times 10^{-3} \,\text{(N/m)/}\text{K}$ \cite{Zeppieri2001}. 
 As shown in Fig.~\ref{fig:marangoni_surfer}, the heated part of the particle boundary is a spherical cap. As defined previously, the coverage factor $\phi$ describes the ratio between the area of the spherical cap and the surface area of the particle, which we set to 0.25 unless otherwise stated. Moreover, the influence of the coverage factor will be investigated in more detail in a later section. For the cases including the effect of surfactant, the additional parameters are $D_\text{s} = 10^{-9} \, \text{m$^2$/s}$ and $\delta = 2.4 \times 10^{3} \, \text{J/mole}$. Note that the value for $\delta$ is identified with $RT$, where $R$ is the universal gas constant \cite{Homsy1984}, using a constant temperature of approximately $300 \, \text{K}$. Due to the linear relation between $\delta$ and $T$, the temperature dependence of $\delta$ is not expected to play a large role and is therefore not included in the model. Finally, we note that all simulations are performed in dimensionless form, but the results are scaled back for easier comparison to the experiments, unless explicitly noted otherwise.

\subsection{Transient simulations} 
\begin{figure}[ht]
	\begin{center}
		\includegraphics[scale=0.6]{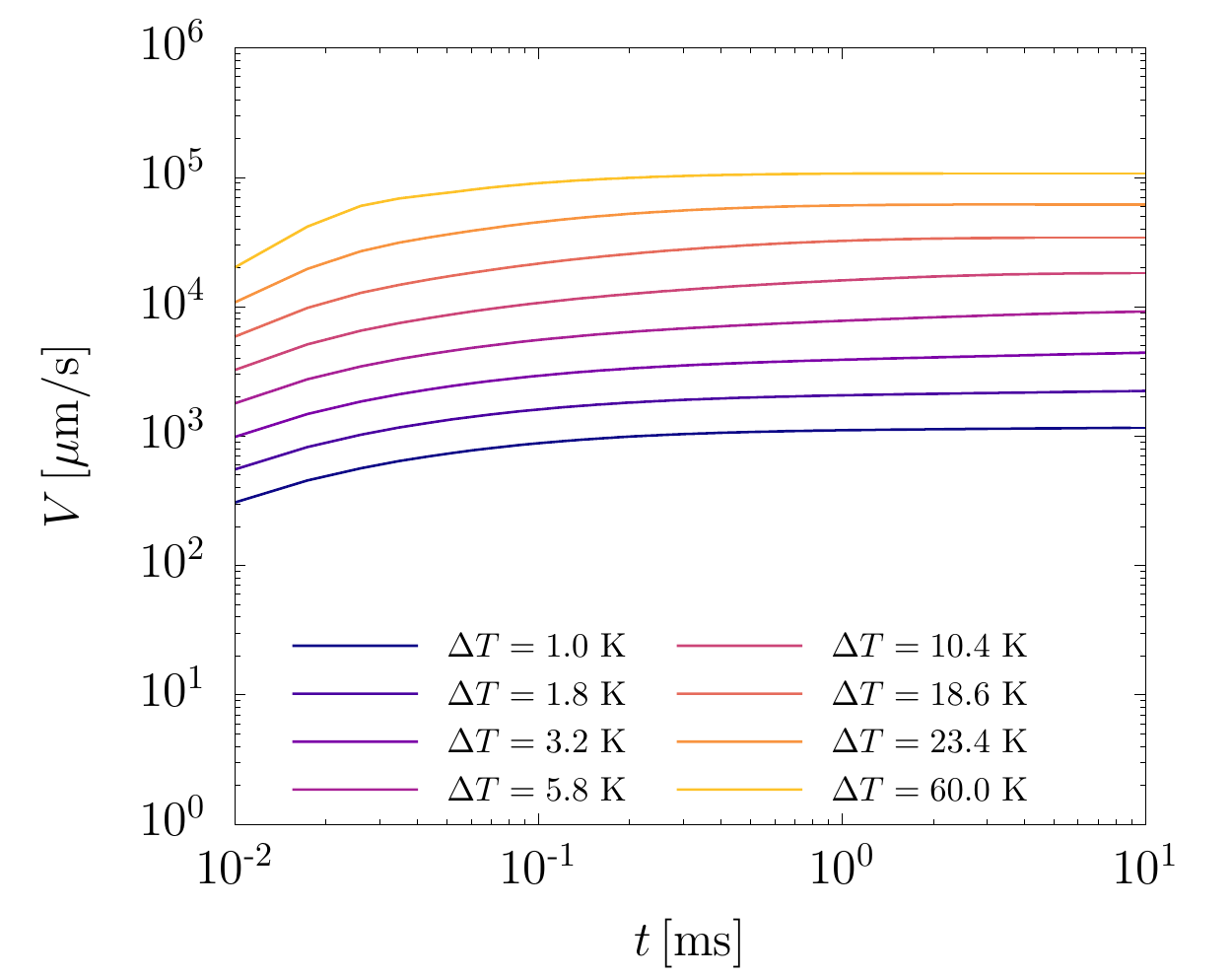}
		\includegraphics[scale=0.6]{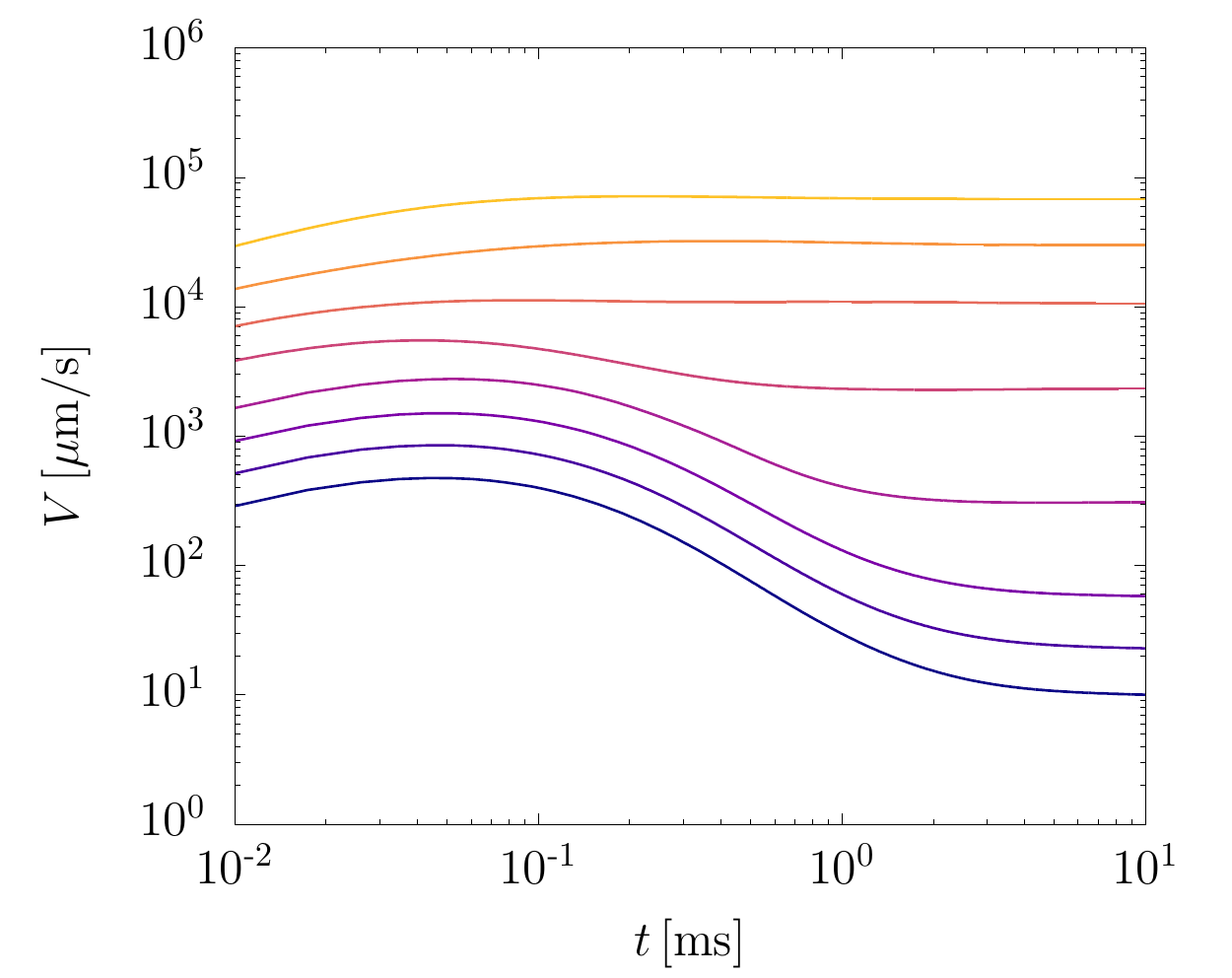}
	\end{center}
	\caption{Transient particle speed $V$ for a clean interface (left), and for an interface with an initial surfactant concentration of $\Gamma_0 = 
	10^{-7} \, \text{mole/m$^2$}$ (right).}
	\label{fig:transient}
\end{figure}

\begin{figure}[ht]
	\begin{center}
		\includegraphics[scale=0.6]{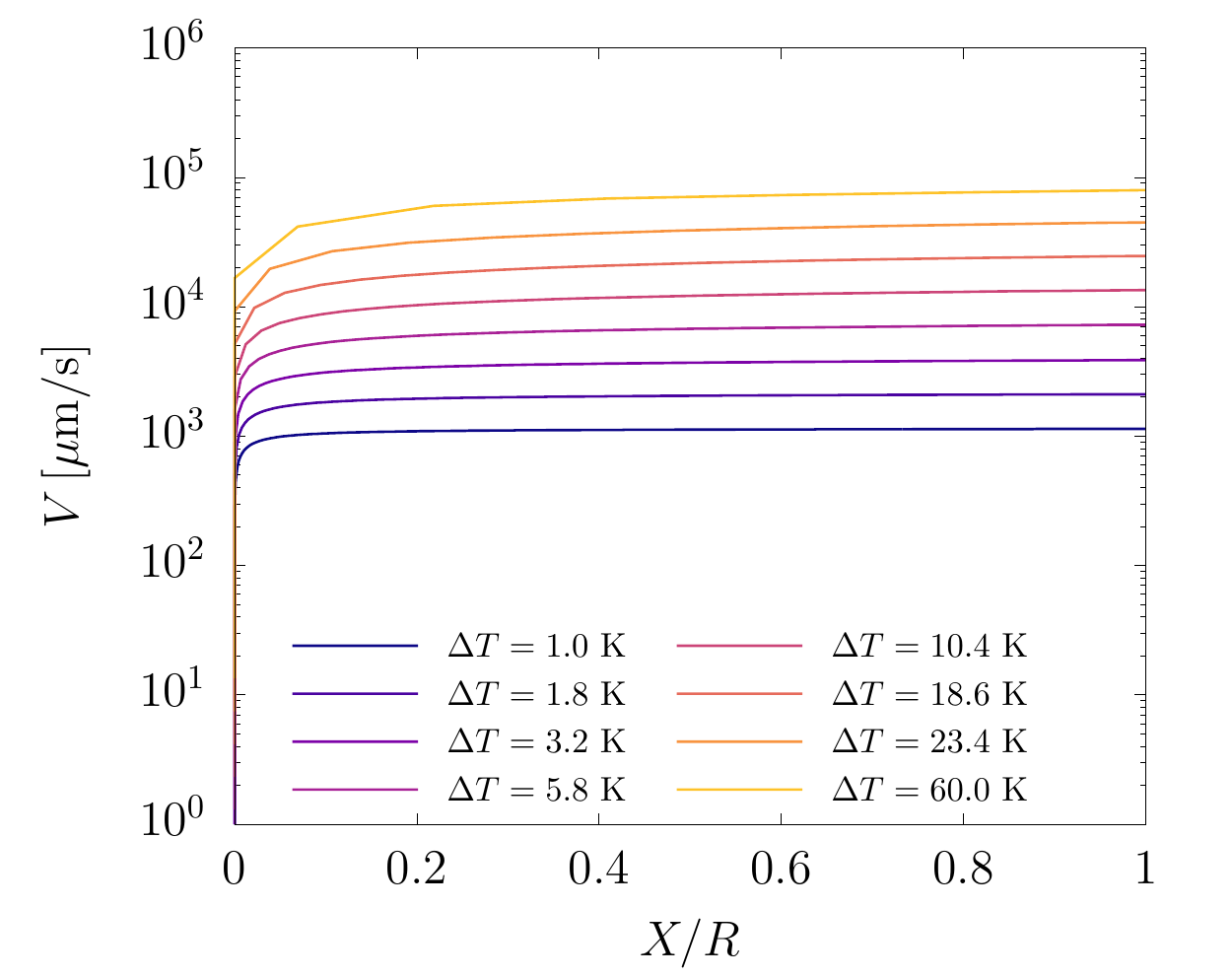}
		\includegraphics[scale=0.6]{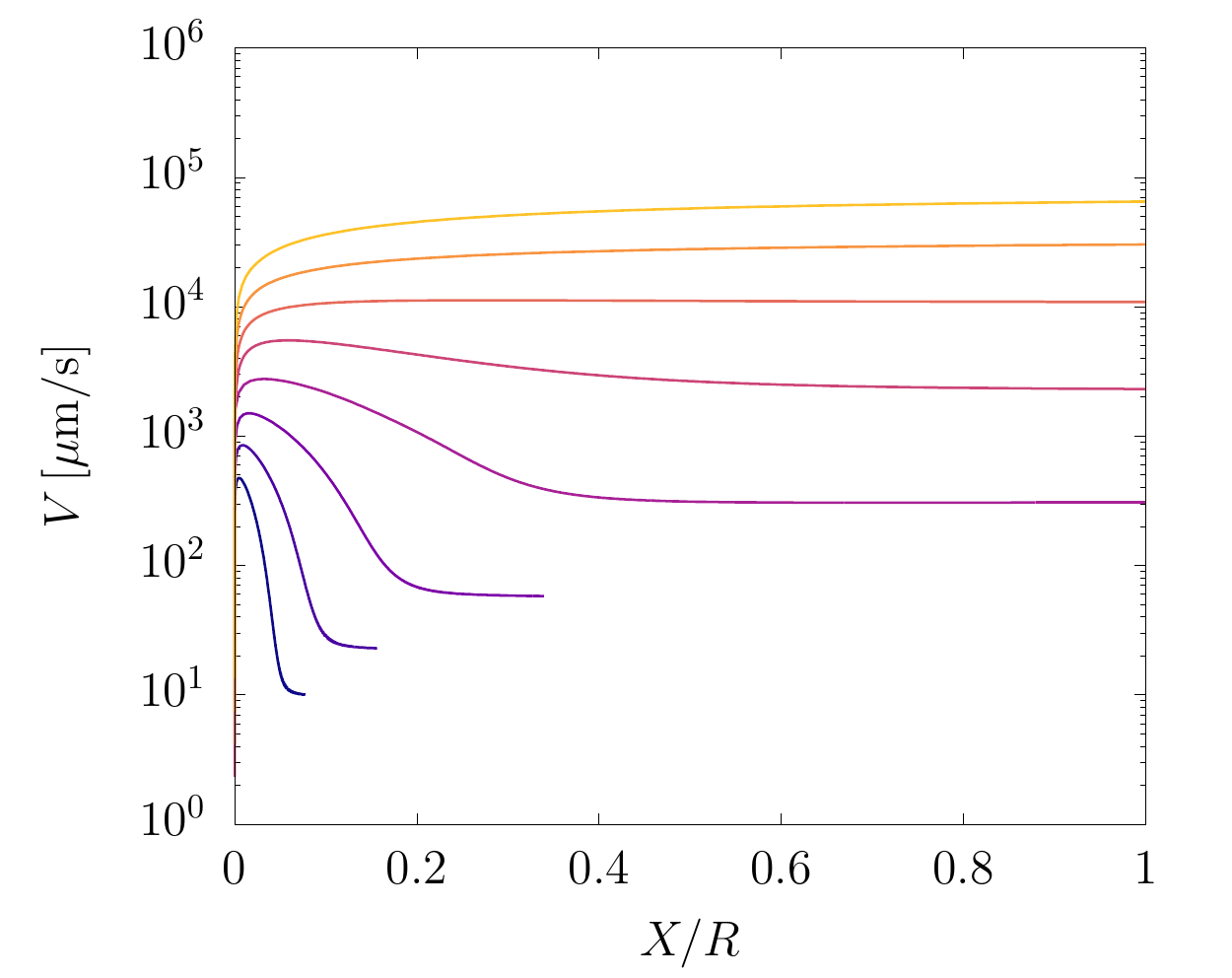}
	\end{center}
	\caption{Transient particle speed $V$ as a function of dimensionless distance traveled for a clean interface (left), and for an interface with an initial surfactant concentration of $\Gamma_0 = 
		10^{-7} \, \text{mole/m$^2$}$ (right).}
	\label{fig:transient_rescaled}
\end{figure}

To investigate the transient behavior of the system, simulations were performed while varying $\Delta T$ between 1 K and 60 K both for a clean interface and an interface where surfactants are present. 
In this section, the coverage factor is 0.25.
The transient particle speed in $x$ direction, denoted by $V$, is shown in Fig.~\eqref{fig:transient} for a clean interface, and for an interface with an initial surfactant concentration of $\Gamma_0 = 10^{-7} \, \text{mole/m$^2$}$. The results indicate that a steady state is reached in about $\unit[10]{ms}$ at lower $\Delta T$, whereas it only takes about $\unit[0.1]{ms}$ at higher $\Delta T$. However, for comparison to the experiments, a more relevant question is over which distance the particle reaches steady state. We therefore show the same data, but now as a function of distance traveled normalized by the particle radius $R$, in Fig.~\ref{fig:transient_rescaled}. The data clearly show that the steady-state velocity is reached well before the particles move a particle radius. Therefore, we can safely investigate the behavior of this system by focusing on the steady-state solutions.

\subsection{Comparison to multipole solution}
\begin{figure}[ht]
	\begin{center}
		\includegraphics[scale=0.12]{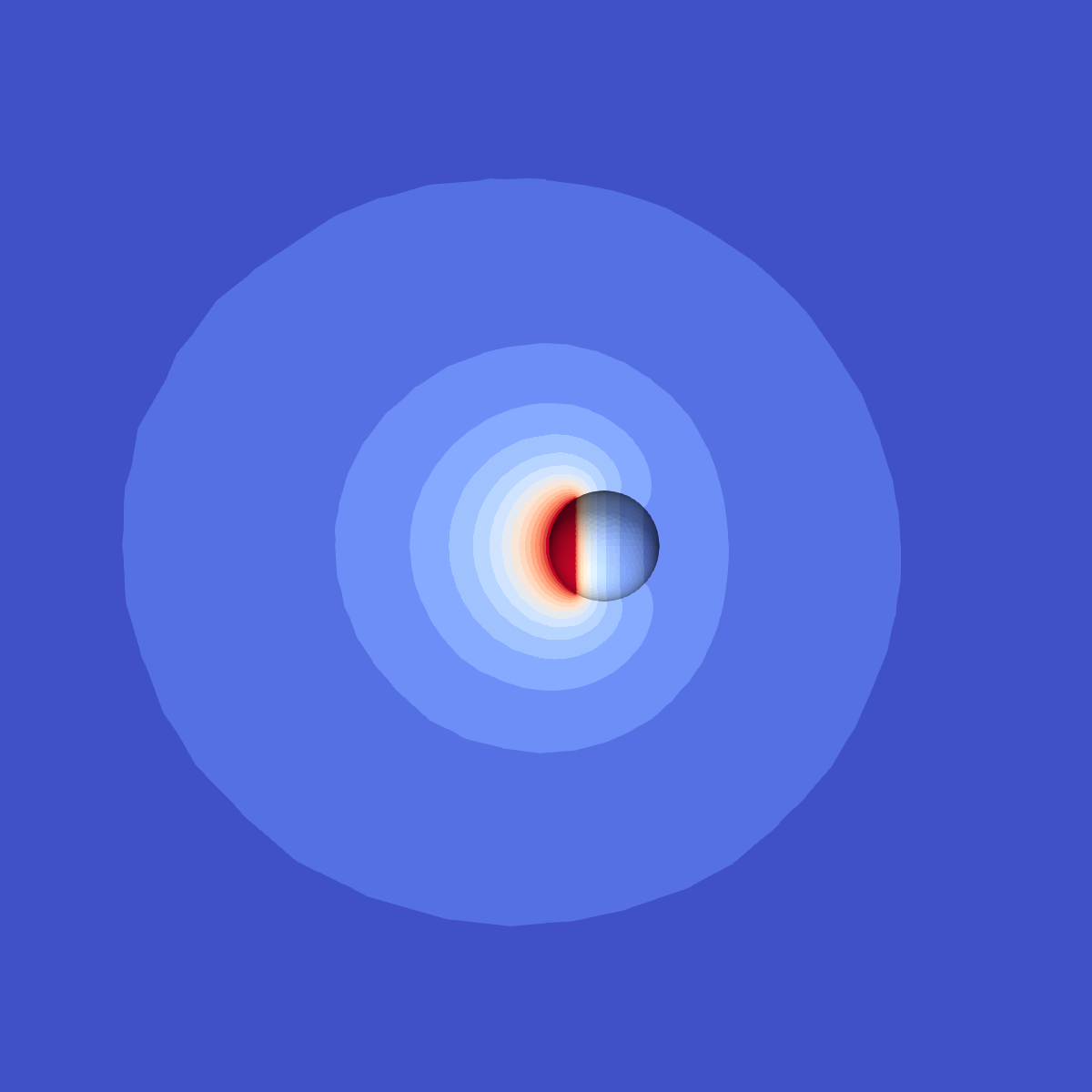} 
		\hspace{0.8cm}
		\includegraphics[scale=0.12]{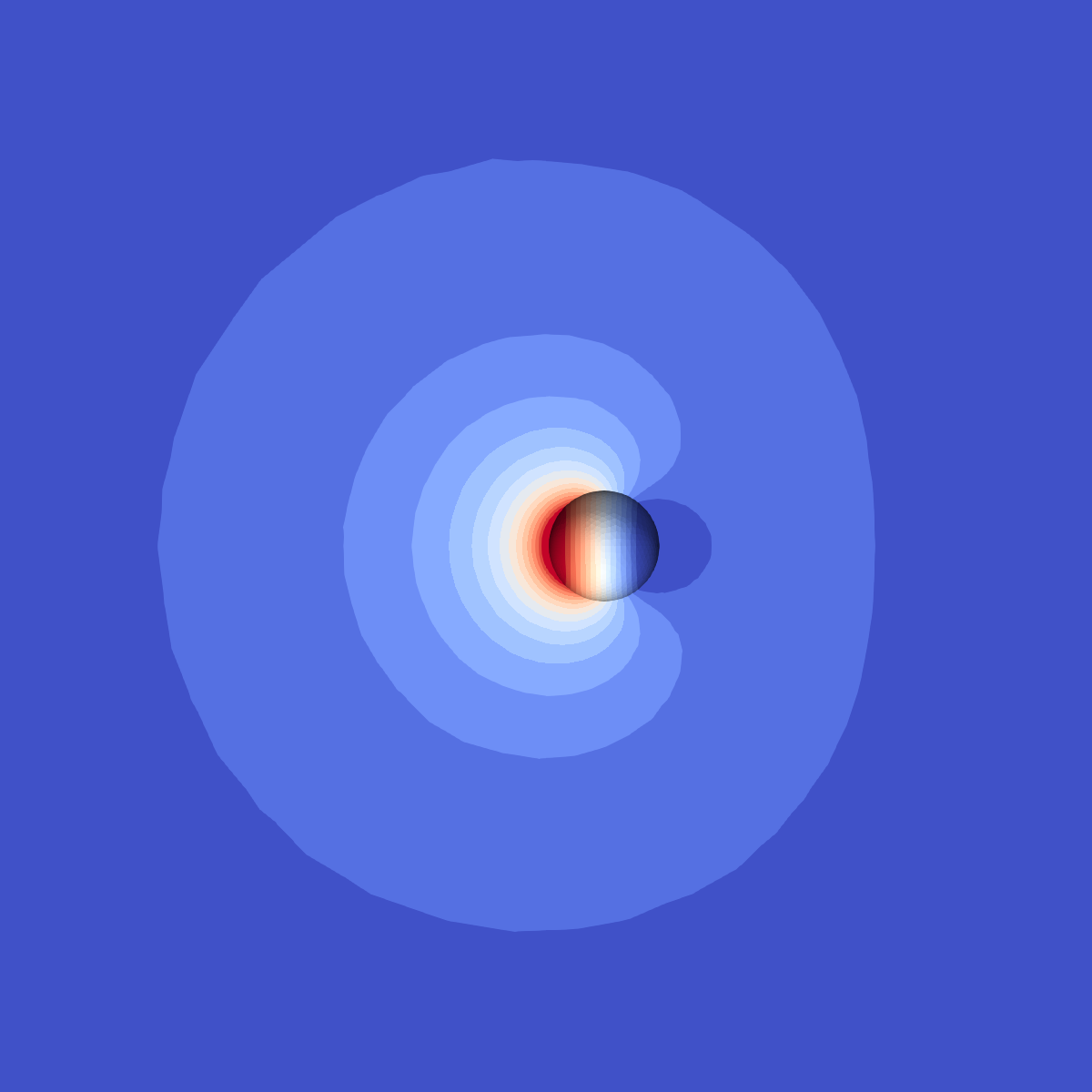}
		\includegraphics[scale=0.15]{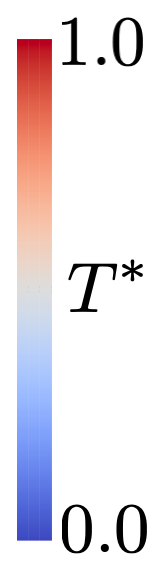}
	\end{center}
	\caption{Comparison of the temperature field as obtained from the simulations (left) and the corresponding temperature field from the multipole expansion (right).}
	\label{fig:temp_sim_mp}
\end{figure}

We compare our numerical solutions to the the analytical solution of W\"{urger} \cite{wurger2014thermally}, which was obtained using a multipole expansion of the temperature field, and which is valid for clean interfaces in the limit of $Pe_{T}=0$. In this approach the temperature field is given by 
\begin{equation}
T(r) = T_0 + a \Delta T_\text{mp} (\frac{1}{r}+\frac{\vec b \cdot \vec 
	r}{r}+\dots), \label{eq:multipole}
\end{equation}
where $\Delta T_\text{mp}$ is the excess temperature in the multipole expansion and $\vec b=-b \vec e_x$ is the temperature dipole vector. Note that $\Delta T_\text{mp}$ is not necessarily equal to $\Delta T$ as we define it. In order to find $\Delta T_\text{mp}$ and the reduced dipole moment $b/a$, we fit Eq.~\eqref{eq:multipole} on the numerical solution by solving a least-squares problem for the case of $\Delta T = 0.001$ K, to ensure convection does not play a role. Integration is performed on half of the domain, centered around the particle, to avoid influence of the boundary conditions. From the solution of the least-squares problem, we obtain $\Delta T_\text{mp}=0.48 \Delta T$ and $b/a=1.39$, for a particle with $\phi=0.25$. A comparison of the temperature field as obtained from the simulations and the corresponding temperature field from the multipole expansion is shown in Fig.~\ref{fig:temp_sim_mp}. Following W\"{u}rger, we can now write:
\begin{equation}
  V=\frac{b}{a}\frac{\beta \Delta T_\text{mp}}{16(\eta_1+\eta_2)}.
\end{equation}

\subsection{Mesh convergence}
\begin{figure}[ht]
	\begin{center}
		\includegraphics[width=0.7\textwidth]{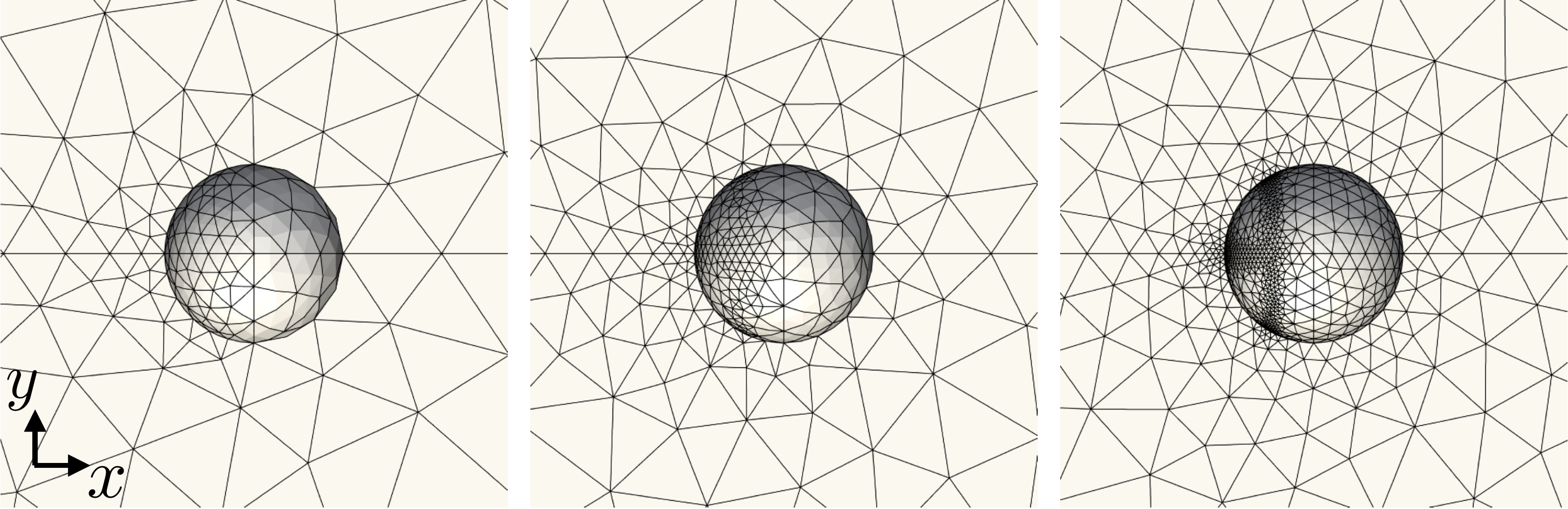}
	\end{center}
	\caption{Close-up of meshes M1 (left), M2 (middle) and M3 (right) as used 
		in the mesh convergence study.}
	\label{fig:refinement}
\end{figure}

\begin{figure}[ht]
	\begin{center}
		\includegraphics[width=0.5\textwidth]{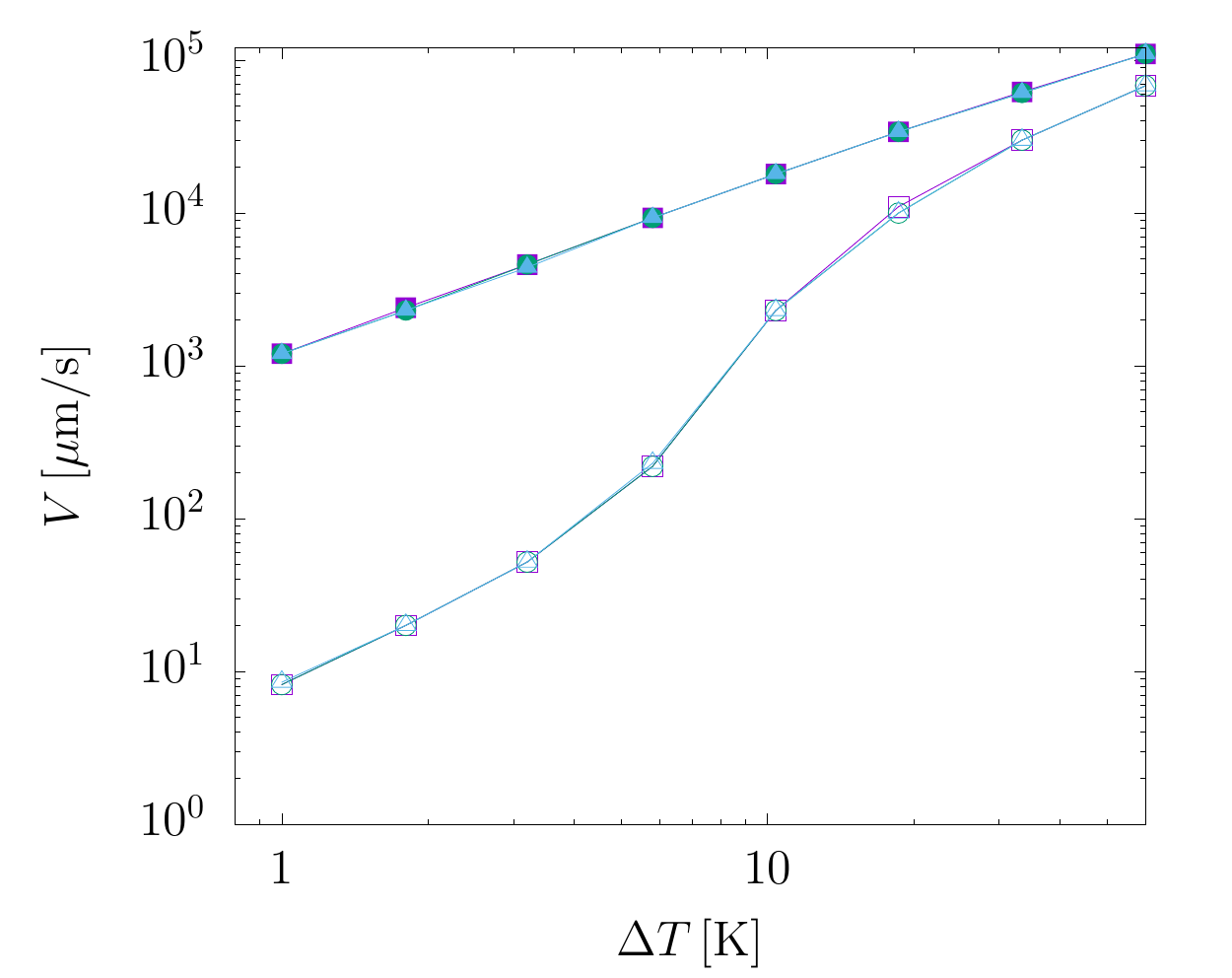}
	\end{center}
	\caption{Steady-state velocity as calculated on the three meshes: M1 
	(purple squares), M2 (blue circles) and M3 (blue triangles). The closed 
	symbols are for a clean 
	interface whereas the open symbols are for
	an interface with an initial surfactant concentration of $\Gamma_0 = 
	10^{-7} \, \text{mole/m$^2$}$.}
	\label{fig:refinement_plot}
\end{figure}

To ensure that the numerical solution is independent of the mesh, we perform the same simulations using three different meshes. A close-up of the mesh around the particle, where the largest gradients are expected to occur, is shown in Fig.~\ref{fig:refinement}. Three meshes are used: a coarse mesh (M1), a medium mesh (M2) and a fine mesh (M3). Note, that the meshes are refined close to the particle, and in regions near the particle where large gradients occur. The steady-state velocities are shown in Fig.~\ref{fig:refinement_plot} as obtained on the three meshes. Almost perfect overlap is obtained, for both the clean interface, and the interface with $\Gamma_0 = \unit[10^{-7}]{mole/m^2}$. To ensure converged solutions for all cases, we performed all simulations on M3.

\subsection{Coverage factor}
\begin{figure}[ht]
	\begin{center}
		\includegraphics[width=0.5\textwidth]{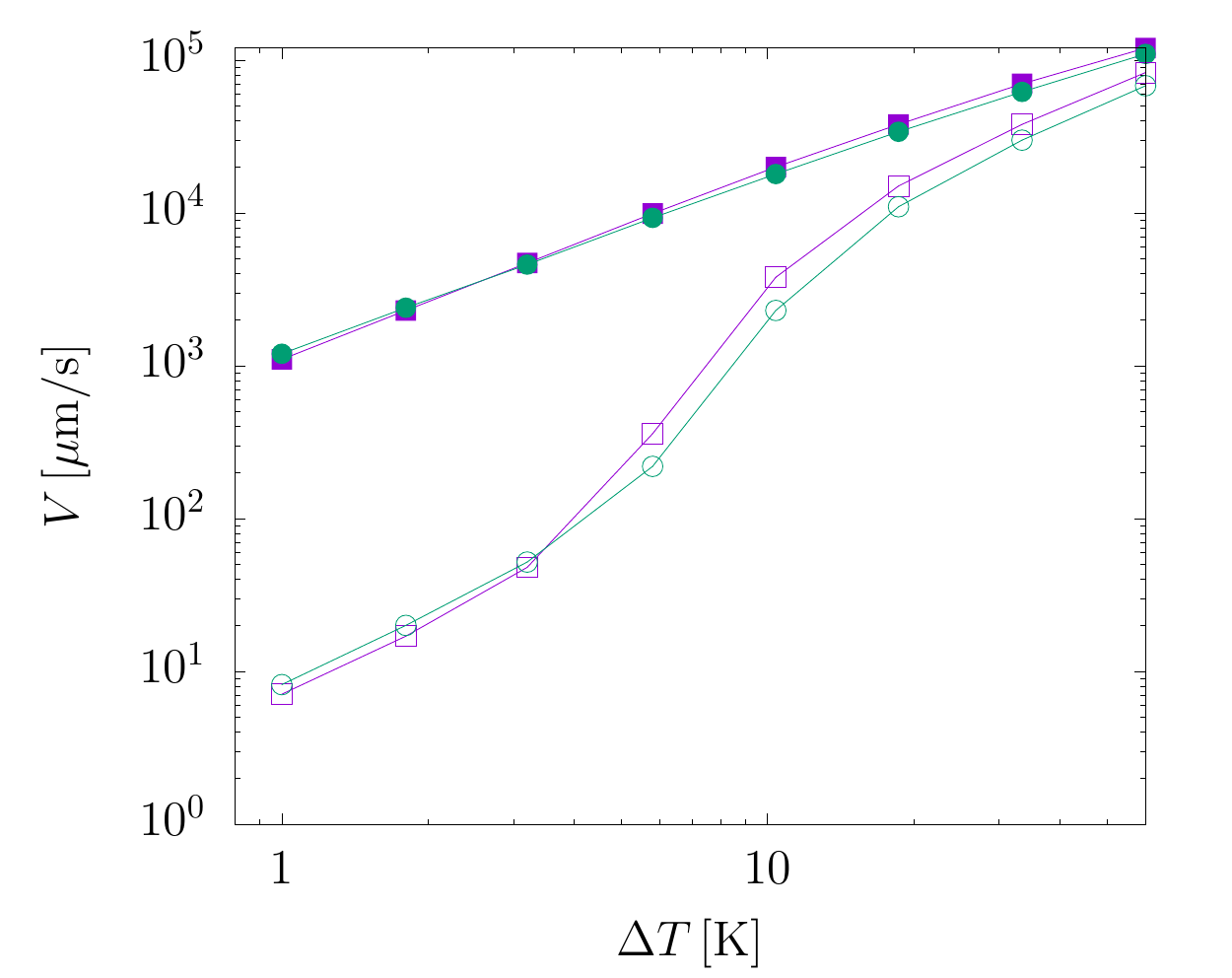}		
	\end{center}
	\caption{Steady-state velocity as calculated for 
	$\phi=0.25$ (purple squares) and $\phi=0.5$ (green circles). The 
	closed symbols are for a clean interface whereas the open symbols are for
	an interface with an initial surfactant concentration of $\Gamma_0 = 
	10^{-7} \, \text{mole/m$^2$}$.}
	\label{fig:coverage}
\end{figure}

We conclude by investigating the influence of the size of the spherical cap, by performing simulations with coverage factors of $\phi=0.25$ and $\phi=0.5$ for a clean interface, as well as for an interface with $\Gamma_0 = 
	10^{-7} \, \text{mole/m$^2$}$. The results are shown in Fig.~\ref{fig:coverage}, and show that the differences between the two cases are small, and well within the accuracy of the experiments. For numerical reasons, simulations with $\phi=0.25$ are more stable, which was therefore used for all results presented in the main manuscript.

\clearpage
\bibliographystyle{plain}
\bibliography{refs}